\documentclass[12pt]{article}
\RequirePackage[colorlinks,citecolor=blue,urlcolor=blue]{hyperref}
\usepackage{amsmath}
\usepackage{graphicx}
\usepackage{enumerate}
\usepackage[authoryear]{natbib}
\usepackage{url} 
\usepackage{amssymb}
\usepackage{bm}
\usepackage{subcaption}
\usepackage{amsfonts}
\usepackage{array}
\usepackage{textcomp}
\usepackage{stfloats}
\usepackage{url}
\usepackage{verbatim}
\usepackage{hyperref} 
\usepackage{booktabs}
\usepackage{nicefrac}    
\usepackage{microtype}   
\usepackage{xcolor}      
\usepackage{multicol}
\usepackage{balance}
\usepackage{notoccite}
\usepackage{algorithmic}
\usepackage[ruled,vlined]{algorithm2e}
\usepackage{makecell}
\usepackage{colortbl}
\usepackage{amsthm}
\usepackage{float}

\newcommand{\blind}{1}

\definecolor{myBlue}{HTML}{2166AC}

\newcommand{\bluesquare}{\textcolor{myBlue}{\rule{1.5ex}{1.5ex}}}
\newcommand{\lightgraysquare}{\textcolor{lightgray}{\rule{1.5ex}{1.5ex}}}

\makeatletter
\newenvironment{algoproc}[1][]
  {\renewcommand{\algorithmcfname}{Procedure}%
   \begin{algorithm}[#1]
   \long\def\@caption##1[##2]##3{%
     \par
     \begingroup\@parboxrestore
     \if@minipage\@setminipage\fi
     \normalsize \@makecaption{\AlCapSty{\AlCapFnt\algorithmcfname}}{\ignorespaces ##3}%
     \par\endgroup
   }}
  {\end{algorithm}}
\makeatother

\begin{document}

\def\spacingset#1{\renewcommand{\baselinestretch}%
{#1}\small\normalsize} \spacingset{1}
\if1\blind
{
  \title{\bf Granger Causality in High-Dimensional Networks of Time Series}
  \author{{\bf{Sipan Aslan and Hernando Ombao}}\vspace{.1cm}\\
    \small{King Abdullah University of Science and Technology (KAUST)} \vspace*{-.1cm} \\
    \small{Statistics Program, CEMSE Divison}}
  \date{}
  \maketitle
} \fi

\if0\blind
{
  \bigskip
  \begin{center}
    {\LARGE\bf Granger Causality in High-Dimensional Networks of Time Series}
\end{center}
  \medskip
} \fi

\bigskip\vspace*{-.5cm}
\begin{abstract}
\noindent A novel and practical approach is introduced for identifying causal connectivity between specific node pairs in high-dimensional networks (HDN) of brain signals. The novelty lies in enabling a conventional Granger causality analysis to work effectively in a high-dimensional setting by combining it with frequency-domain-based spectral dynamic principal component analysis (sDPCA). The proposed procedure begins with the estimation of a low-dimensional representation of the other nodes in the network utilizing sDPCA. The resulting scores can then be extracted from the nodes of interest, thus removing the confounding effect of other nodes within the HDN. Accordingly, causal interactions can be dissected between nodes that are isolated from the effects of the network. Simulations confirm that the approach reliably retrieves causal links, even in highly interactive networks. The method also performs well on multichannel EEG data, and findings align with earlier findings in the literature.

\end{abstract}

\noindent%
\vfill

\newpage
\spacingset{1.5} 

\section{Introduction}
Identifying effective connectivity in high-dimensional networks (HDN) remains a demanding problem and has been addressed by several studies \citep[see, e.g.,][]{kalisch2014causal,wang2016modeling,ting2016estimation,ting2017estimating,wang2019exploratory,zarghami2020dynamic,siggiridou2021dimension,shojaie2022granger}. Moreover, it is difficult to infer or interpret interactions between a pair of nodes while taking into account potential interactions with all other nodes in the background \citep{wang2020high,basu2024high}. One particular challenge arises when the objective is simply to identify the causal interaction between specifically interested pairs of nodes or pairs of subsets of nodes/channels within an extensive network of interacting nodes. Attempting to model an entire HDN for such a specific objective will require fitting a highly parameterized model with potentially hundreds of parameters, which can lead to over-parameterization of the task and interpretation difficulties. 

In fact, utilizing conventional bivariate vector autoregressive models (VAR) can be a practical and manageable way to identify pairwise causal interactions in high-dimensional networks as long as the potential confounding effect due to the high-dimensional background nodes is addressed. Adopting this strategy, we propose a procedure that uses conventional Granger causality (GC) testing to detect causal connections between any two nodes while accounting for the global background activity in a low-dimensional representation. Such a representation is constructed through spectral dynamic principal component analysis (sDPCA) \citep{brillinger1969canonical,brillinger1981time}, which extends the principal component analysis to incorporate temporal dependence. By working with the spectral density matrix, directly linked to all auto and cross-covariances, sDPCA preserves frequency-specific features critical for neural data analysis.

The sDPCA enhances the PCA by incorporating temporal dependency within the data, thereby reducing the dimensionality of time series in the frequency domain while retaining temporal information. In other words, while the conventional PCA is based on the eigendecomposition at the zero-lag (or contemporaneous) covariance matrix, sDPCA is based on the eigendecomposition of the spectral density matrix estimated for lagged auto-covariances at all frequencies. Since there is a 1-1 relationship between the spectral matrix (at all frequencies) and the covariance matrix (at all lags), this approach captures lagged cross-correlations amongst signals. From a Bayesian paradigm, the generalized probabilistic PCA developed by \cite{gu2020generalized} provides a dimension-reduction mechanism that accounts for serial and spatial correlation in multivariate signals. 

Various methodologies have been derived from frequency domain PCA. For instance, \cite{stoffer1993spectral, stoffer1999detecting,stoffer2000spectral} developed a spectral envelope using dynamic PCA to identify common signals in multiple categorical and real-valued time series, \cite{hormann2015dynamic} proposed a variant of dynamic PCA for analyzing functional time-series data. As demonstrated by \cite{ombao2006time, wang2019exploratory,ombao2022spectral}, principal component analysis in the frequency domain offers practical benefits and is highly effective in summarizing high-dimensional signals by capturing their main modes of variation. On the other hand, given the multivariate nature of the objectives outlined herein, it becomes relevant to inquire about the applicability of the state-of-the-art techniques of multivariate time series analysis. For example, Factor-augmented VAR (FAVAR) models \citep{bernanke2005measuring, lin2020regularized} reduce dimensionality by extracting a handful of static factors from large panels and then placing these factors into a more manageable VAR system. This strategy centers on global modeling and treating the extracted factors as regressors to explain multiple target variables; however, it is not designated for cases where attention is on a single pair of nodes within a high-dimensional network. Dynamic factor models (DFMs) \citep{forni2000generalized,forni2015dynamic,NBERw11467,barigozzi2017network} similarly rely on a few latent components, focusing on decomposing each series into common and idiosyncratic components. DFM allows more detailed lagged structures than static factor approaches, but often seeks to explain broad data variance or perform large-scale forecasts. Bayesian mixture network models can likewise account for cross-graph heterogeneity \citep{yin2022finite}. Conditional Granger causality (CGC) \citep{geweke1984measures} instead tests whether adding one predictor refines forecasts of another, holding all others fixed. Although CGC reveals causal associations, it can become complicated in high dimensions when many variables must be jointly conditioned. While there are some parallels between the proposed approach and that of others, each method is characterized by a distinct emphasis. The proposed approach diverges by isolating specific node pairs. It accounts for background effects by spectral components and then applies GC testing on a simplified bivariate system. This diverges from CGC's exhaustive conditioning on many channels, from FAVAR's global factor plus VAR, and dynamic factor models’ emphasis on large-scale shared patterns. The approach also provides feasibility in large networks, where a smaller factor-driven representation is sufficient to identify causal interactions.

The proposed approach distinguishes itself from existing approaches by focusing on frequency-domain dimensionality reduction combined with targeted pairwise causality analysis. The theoretical formulations of CGC, FAVAR, and DFMs have different aims – identifying influences through conditioning, incorporating large information sets into VARs, and explaining common variance in high dimensions, respectively – and each comes with certain assumptions about the data-generating process. The proposed approach does not compete with these state-of-the-art methods on their own terms; rather, it occupies a complementary cavity. It assumes that in many high-dimensional systems (e.g., brain recordings), one can extract the key oscillatory components and concentrate on their interactions, thereby enabling meaningful causal understanding without having to model every other interaction. Empirically, this translates to a convenient method that is sensitive to the features of brain monitoring data – handling many channels and frequency-specific processes – while building on the proven strengths of factor modeling and Granger causality. Given the objectives of our approach, it is important to note that GC combined with sDPCA does not propose a universal substitute for full network modeling. Instead, it is a domain-orientated alternative that emphasizes interpretability and feasibility in the analysis of high-dimensional network time series. This allows for targeted analysis, whether at the level of individual node pairs or among a tractable number of node clusters, without being constrained by finitely many interactions. In what follows, Section 2 presents the proposed approach for an HDN setting, Section 3 describes simulation setups and results, and Section 4 features an application to EEG recordings, showing how the method handles brain signals.


\section{Methodology}
\subsection{Inference in high-dimensional networks of time series}
Define $V  = \{Y^*_t, X^*_t, Z_{1,t}, \ldots, Z_{n-2,t}\}$ to be a high dimensional network of signals with possibly $n \geq 256$ nodes. As we consider in this paper, this network may represent the activity of a brain through EEG recordings in which each component (i.e., node or channel) is actually a time series. In this notation of network $V$, $Y^*$ and $X^*$ denote any two nodes of interest. For problems whose solutions are sought component-wise, we set our goal to explore the effective connectivity between any node pairs in a network rather than the associativity of the whole network. Node pairs are subject to the influence of other nodes due to the network's relational structure. This means that to dissect the respective effective connection between any two nodes of interest; they need to be freed from the influences of other nodes. Thereby, the interference-free counterparts of the nodes of interest can be obtained in the following manner:
\[ \hat{Y}_t = Y^*_t - \hat{Y}^*_t|f_z(Z_{1,t}, \ldots, Z_{n-2,t}), \]
\[ \hat{X}_t = X^*_t - \hat{X}^*_t|f_z(Z_{1,t}, \ldots, Z_{n-2,t}), \]

where the $f_z(.)$ is a function that encompasses the dynamic variability of excluded nodes. Practically, the function $f_z(.)$ acts as a filter and can take any form as long as it provides the net summary discharge of the nodes excluded. In the scope of this paper, the function $f_z(.)$ is intended to provide low-dimensional summaries of excluded nodes. To this end, we utilize the spectral-domain dynamic PCA implemented in the \texttt{R} package \texttt{freqdom} made available by \cite{hormann2015dynamic}. Spectral PCA originally proposed by \cite{brillinger1981time}, enables us to obtain principal component scores that concisely encapsulate variability in multi-dimensional time series, specifically with regard to their serial dependence \citep[see, e.g., ]{shumway2000time,ombao2022spectral}. Hence, we propose the filter $f_z(\cdot)$ over the excluded nodes to yield a vector of summary indicators,
\[
  f_z(Z_{1,t}, \ldots, Z_{n-2,t}) = \mathbf{p}_t = \bigl( dpc_{1}(t), dpc_{2}(t), \ldots, dpc_{k}(t)\bigr),
\]
where $dpc_{i}(t)$ denotes the score of the $i$-th dynamic principal component obtained through sDPCA. One may optionally augment \(\mathbf{p}_t\) with higher-order transformations of the principal-component scores (for instance, cross-products \(pc_i(t) \times pc_j(t)\) or polynomial expansions) if a more general functional form is required to capture nonlinearities or interactions among these components.

In general, the conditional mean of $Y^*_t$ and $X^*_t$ given these indicators can be modeled as:
\[
  \hat{E}\bigl(Y^*_t \big\vert \mathbf{p}_t\bigr)
  = G_{Y^*}\bigl(\mathbf{p}_t; \boldsymbol{\theta}_{Y^*}\bigr),
\]
\[
  \hat{E}\bigl(X^*_t \big\vert \mathbf{p}_t\bigr)
  = G_{X^*}\bigl(\mathbf{p}_t; \boldsymbol{\theta}_{X^*}\bigr),
\]
where $G_{Y^*}(\cdot;\boldsymbol{\theta}_{Y^*})$ and $G_{X^*}(\cdot;\boldsymbol{\theta}_{X^*})$ are functions (e.g.\ linear, nonlinear, or another parametric/nonparametric regression form) parameterized by $\boldsymbol{\theta}_{Y^*}$ and $\boldsymbol{\theta}_{X^*}$.

Considering an option where the terms are linearly combined and have first-order interactions, the representation can be written as follows:
\[
  G_{Y^*}\bigl(\mathbf{p}_t; \boldsymbol{\theta}_{Y^*}\bigr)
  = \alpha_0 + \sum_{i=1}^{q} \alpha_i dpc_{i}(t)
  + \sum_{i=1}^{q} \sum_{j \neq i}^{q}  \alpha_{ij} \bigl[ dpc_i(t) \times dpc_j(t)\bigr],
\]
Here, $q$ is typically small relative to the original dimension,
so that $\{\alpha_0, \alpha_i, \alpha_{ij}\}$ remain tractable to estimate in practice.

\paragraph{Spectral DPCA} The frequency domain dynamic PCA decomposes a multivariate time series into uncorrelated components in the frequency domain, maximizing the long-run variance explained. Unlike classical PCA, which produces components uncorrelated in space, sDPCA focuses on the spectral density of the data, producing components uncorrelated in time.

Assume a multivariate time series $\mathbf{Z}$ of dimension $T \times n$, where $T$ is the number of time points, and $n$ is the number of variables; this can be considered as a network of time series. The first step in sDPCA involves estimating the spectral density matrix $F(\omega)$ from the time series. The cross-spectral density for $\mathbf{Z}$ can be defined as:

\[
F_{\mathbf{Z}}(\omega) = \sum_{h \in \mathbb{Z}} \text{Cov}(\mathbf{Z}_{h}, \mathbf{Z}_{0})  \mathrm{exp}{(-2\pi i h \omega)},
\]

where $\text{Cov}(\mathbf{Z}_{h}, \mathbf{Z}_{0})$ is the auto-covariance matrix at lag $h$. The empirical cross-spectral density can be estimated using a windowed version of the lagged auto-covariance matrices:

\[
\hat{F}_{\mathbf{Z}}(\omega) = \sum_{|h| \leq q} w\left(\frac{|h|}{q}\right) \hat{C}_{\mathbf{Z}}(h)  \mathrm{exp}{(-2\pi i h \omega)},
\]

where $w(\cdot)$ is a window function (i.e., kernel), $q$ is the window size, and $\hat{C}_{\mathbf{Z}}(h)$ is the empirical lagged auto-covariance given by:

\[
\hat{C}_{\mathbf{Z}}(h) = \frac{1}{T} \sum_{t=1}^{T-|h|} (\mathbf{Z}_{t+|h|} - \bar{\mathbf{Z}})(\mathbf{Z}_t - \bar{\mathbf{Z}})',
\]

for $h \geq 0$, and

\[
\hat{C}_{\mathbf{Z}}(h) = \hat{C}_{\mathbf{Z}}(-h)' = \frac{1}{T} \sum_{t=1}^{T + h} (\mathbf{Z}_{t} - \bar{\mathbf{Z}})(\mathbf{Z}_{t - h} - \bar{\mathbf{Z}})'.
\]

for $h < 0$.

Once the spectral density matrix $F(\omega)$ is obtained, the dynamic principal component filters $\phi_k^{(j)}$ are computed, where $k \in [-q,q] \subset \mathbb{Z}$ and the index $j$ is referring to the $j$-th largest dynamic eigenvalue. The filters are derived as the Fourier coefficients of the dynamic eigenvectors $\varphi_j(\omega)$ of the spectral density matrix $F(\omega)$:

\[
\phi_k^{(j)} = \frac{1}{2\pi} \int_{-\pi}^{\pi} \varphi_j(\omega) e^{-i k \omega} d\omega.
\]

By construction, sDPCA typically employs two-sided filters when forming the scores. This is seen in
\begin{equation*}
\label{eqn:dpc-twosided}
dpc_{t}^{(j)}  = \sum_{k=-q}^{q} \bigl\{\phi_{k}^{(j)}\bigr\}^{\!\top} \mathbf{Z}_{t-k},
\end{equation*}
where $\{\phi_{k}^{(j)}\}$ are the Fourier-based filter coefficients arising from the inverse transform of the $j$-th dynamic eigenvector. The index $k$ runs from $-q$ to $q$, reflecting a non-causal (two-sided) convolution that uses both past and future data relative to time $t$. This formulation can yield improved spectral estimates, though it assumes the entire data record is available.

In real-time or causal applications, a one-sided variant can be adopted:
\begin{equation*}
\label{eqn:dpc-onesided}
dpc_{t}^{(j)}  = \sum_{k=0}^{q} \bigl\{\phi_{k}^{(j)}\bigr\}^{\!\top} \mathbf{Z}_{t-k},
\end{equation*}
which omits “future” values $\mathbf{Z}_{t+k}$ for $k>0$. This causal filter operates only on current and lagged observations, making it suitable for sequential or forecasting tasks. The choice between two-sided and one-sided implementations depends on whether retrospective or real-time usage is intended, as well as on the degree to which one wishes to exploit possible gains from including both past and future observations.






The proportion of variance explained by the $j$-th dynamic principal component is quantified as:

\[
v_j = \frac{\int_{-\pi}^{\pi} \lambda_j(\omega) d\omega}{\int_{-\pi}^{\pi} \text{tr}(F(\omega)) d\omega},
\]

where $\lambda_j(\omega)$ is the $j$-th dynamic eigenvalue of the spectral density matrix $F(\omega)$. This ratio provides a measure of how much of the total variance in the time series is captured by each dynamic principal component.

In summary, the sDPCA method involves estimating the spectral density matrix from the time series data using windowed empirical lagged auto-covariances and Fourier transform, computing dynamic principal component filters by performing eigendecomposition on the spectral density matrix and applying the inverse Fourier transform to the dynamic eigenvectors, obtaining dynamic principal component scores by filtering the time series with the computed filters, and quantifying the proportion of variance explained by each dynamic principal component. These steps collectively facilitate the decomposition of multivariate time series into components that are uncorrelated in time and maximize the long-run variance explained.

\subsection{Granger Causality}
Granger Causality (GC) is a statistical concept used to determine whether one time series can predict another. The concept, introduced by \cite{granger1969investigating}, is based on the principle that if the prediction of a time series \(Y_t\) can be improved by incorporating past values of another time series \(X_t\), then \(X_t\) is said to "Granger-cause" \(Y_t\).

The Granger causality analysis involves the following steps. First, consider two time series \(X_t\) and \(Y_t\). To test if \(X_t\) Granger-causes \(Y_t\), we compare the following two models:

\noindent{- Unrestricted Model:}
   \[
   Y_t = \alpha_0 + \sum_{i=1}^{p} \alpha_i Y_{t-i} + \sum_{j=1}^{q} \beta_j X_{t-j} + \epsilon_t
   \]
- Restricted Model:
   \[
   Y_t = \gamma_0 + \sum_{i=1}^{p} \gamma_i Y_{t-i} + \eta_t
   \]
Here, \(p\) and \(q\) are the maximum lags considered, \(\alpha_0\) and \(\gamma_0\) are constants, \(\alpha_i\) and \(\gamma_i\) are coefficients of the lagged \(Y_t\), \(\beta_j\) are coefficients of the lagged \(X_t\), and \(\epsilon_t\) and \(\eta_t\) are error terms.

To determine if \(X_t\) Granger-causes \(Y_t\), we test the null hypothesis:
   \[
   H_0: \beta_1 = \beta_2 = \cdots = \beta_q = 0
   \]
against the alternative hypothesis:
   \[
   H_1: \text{At least one } \beta_j \neq 0 \text{ for } j=1,2,\ldots,q
   \]

The test statistic is based on the F-distribution and is computed as:
   \[
   F = \frac{(RSS_R - RSS_U) / q}{RSS_U / (T - p - q - 1)},
   \]
where \(RSS_R\) and \(RSS_U\) are the residual sum of squares of the restricted and unrestricted models, respectively, and \(T\) is the number of observations. If the computed F-statistic is greater than the critical value from the F-distribution, we reject the null hypothesis, concluding that \(X_t\) Granger-causes \(Y_t\).

Granger causality is widely used in fields such as economics, neuroscience, and environmental sciences. In neuroscience, it helps to understand the directional interactions between brain regions by analyzing EEG or fMRI data. However, conventional GC assumes that error terms are normally distributed and homoscedastic, which is often violated in practice, especially in EEG time series that exhibit heteroscedasticity and fat-tailed distributions. These violations can bias results.
 
Unobserved confounders can also complicate GC analysis. If external variables influencing both time series are omitted, the test might falsely identify causal relationships. This issue is particularly acute in complex systems with many interacting variables.

\subsection{Proposed Approach}

In high-dimensional networks of time series, analyzing causal connectivity between specific pairs of signals can be both challenging and computationally intensive due to the complexity of modeling interactions across all nodes. The approach to be followed would be to model the whole material if there were no high dimensionality problems. However, modeling under high dimensionality requires fitting highly parameterized models to the entire network, leading to difficult-to-interpret estimates and an increased risk of overfitting. To overcome these challenges, we propose a practical and easily implementable method that focuses on isolating the nodes/channels of interest by effectively reducing the dimensionality of the network. Our approach utilizes spectral dynamic principal component analysis to capture the essential variability of the background network in a low-dimensional representation. By partialling out the influence of other nodes through sDPCA, we can isolate the COI and accurately assess the causal connectivity between them using conventional Granger causality framework. This method simplifies the analysis by avoiding the need to model the entire high-dimensional network and enhances interpretability when the primary interest is in the causal relationship between specific signals. The proposed approach is given in the \textbf{Procedure} outline below.

~

{
\setlength{\parskip}{0pt}%
\renewcommand{\baselinestretch}{0.9}\selectfont   

\begin{algoproc}[H]
\SetAlFnt{\normalsize}
\SetAlCapFnt{\normalsize}
\SetAlCapNameFnt{\normalsize}
\SetAlgoInsideSkip{3pt}
\SetAlgoSkip{3pt}
\setlength{\algomargin}{1em}  

\caption{Identifying Causal Connectivity in HDN}
\label{proc}

~

\KwIn{High-dimensional Networks of Signals \( V = \{Y^*_t, X^*_t, Z_{1,t}, \ldots, Z_{n-2,t}\} \)}
\KwOut{Determination of causal relationship between \( Y^*_t \) and \( X^*_t \)}

~

\textbf{1. Identify Nodes of Interest:}
NOI:\( \{ Y^*_t, X^*_t \} \). 
In practice, each node can also be a cluster of nodes represented by sDPCA scores.

\textbf{2. Apply sDPCA:}
\begin{enumerate}[i]
\item Estimate spectral density matrix \( F_{\mathbf{Z}}(\omega) \) for \( \mathbf{Z}_t \).
\item Compute dynamic eigenvectors \( \varphi_j(\omega)\) and eigenvalues \(\lambda_j(\omega)\).
\item Derive dynamic principal component filters \( \phi_k^{(j)}\).
\item Calculate dynamic principal component scores \(\text{dpc}_t^{(j)}\).
\end{enumerate}

\textbf{3. Partial Out Influence of Other Nodes:}
\[Y_t = Y^*_t - \hat{E}\bigl(Y^*_t \big\vert \mathbf{p}_t\bigr) \] 
\[X_t = X^*_t - \hat{E}\bigl(X^*_t \big\vert \mathbf{p}_t\bigr) \]

\textbf{4. Fit Restricted/Unrestricted Models:}
\begin{enumerate}[i]
\item \(\displaystyle Y_t = \alpha_0 + \sum_{i=1}^{p} \alpha_i Y_{t-i} + \sum_{j=1}^{q} \beta_j X_{t-j} + \epsilon_t \).
\item \(\displaystyle Y_t = \gamma_0 + \sum_{i=1}^{p} \gamma_i Y_{t-i} + \eta_t \).
\end{enumerate}

\textbf{5. Testing Granger Causality:}
\begin{enumerate}[i]
\item Test \(H_0: \beta_j = 0\) for all \(j\).
\item Compute F-statistic \(F = \frac{(RSS_R - RSS_U)/q}{RSS_U/(T - K)}\).
\item Reject \(H_0\) if \(F\) exceeds the critical value \( F_{q, T-K}\).
\end{enumerate}

\end{algoproc}

} 


\section{Simulation study}

This section presents a framework for evaluating whether Granger causality analysis, combined with sDPCA-based removal of confounding effects, can recover underlying causal associations in a set of EEG-like time series networks. The main objective is to demonstrate how this approach addresses linear, nonlinear, and causative external influences in high-dimensional settings. The third scenario is evaluated in a comparative framework. In this context, GC analysis with conventional PCA, DFM-based approaches by \cite{barigozzi2024fnets}, and sparse vector autoregressive (VAR) estimation through parameter penalization techniques by \citep{nicholson2017bigvartoolsmodelingsparse,nicholson2020high} can be summoned up to obtain GC associations in large networks. These methods are made available in the \texttt{R} environment. Although the DFM estimation provided by the \texttt{fnets} package and the penalized VAR estimations provided by the \texttt{BigVAR} package are not directly designed for GC analysis, they can be regarded as alternatives since the estimated sparse coefficient matrices can signify GC links among variables. We have also considered alternatives where tests for GC detection can be performed with these techniques. All simulations incorporate designed connections among specific channels, while outer nodes of networks introduce confounding influences to obscure them. 

Suppose the total size of the generated networks is $N + N_{external}$ in which signals are labeled as \(X_1, \ldots, X_{N/2}, Y_1, \ldots, Y_{N/2}, Z_1, \ldots, Z_{N_{\mathrm{ext}}}\). The goal is to explore how well one can detect the causal connection \(X_i \to Y_i\) after these pairs are affected by additional channels $\{Z_j\}$. Each $X^*_i$ is produced by mixing latent oscillatory components that reflect distinct frequency bands commonly used to approximate EEG activity. These bands—delta ($0.5$--$4$ Hz), theta ($4$--$8$ Hz), alpha ($8$--$12$ Hz), beta ($12$--$30$ Hz), and gamma ($>$ $30$ Hz)—are treated as AR(2) processes of the form \( \nu(t) = \psi_{1\nu} \nu(t-1) + \psi_{2\nu} \nu(t-2) + \eta_{\nu}(t), \) where \(\psi_{1\nu} = 2 r_{\nu}\cos(\varrho_{\nu})\), \(\psi_{2\nu} = - r_{\nu}^2\), and \( \eta_{\nu}(t) \sim \mathrm{WN}(0,\sigma^2_\nu) \). The resulting latent components, for \(\nu \in \{\delta,\theta,\alpha,\beta,\gamma\}\), allow each $\nu$ to maintain a stationary and nearly resonant character. A typical $X_i$ can be written as
\[
X_i(t)
= a_i \delta(t) + b_i \theta(t) + c_i \alpha(t) + d_i \beta(t) + e_i \gamma(t) + \varepsilon_i(t),
\]
where \(\{a_i,b_i,c_i,d_i,e_i\}\) are channel-specific weights and \( \varepsilon_i \sim \mathrm{WN}(0,\sigma^2_\varepsilon)\). Its paired $Y^*_i$ is constructed by applying an AR model to $X^*_i$:
\[
Y_i(t)
= \sum_{k=1}^{p} \psi_k  X_i(t-k) + \xi_i(t),  \text{ where } \xi_i(t) \sim \mathrm{WN}(0,\sigma^2_\xi)
\]
with $p=2$ and stationarity-providing coefficients $\psi_k \in \pm 1.5$ in this study. This pairing establishes underlying cause-and-effect association \(X_i \to Y_i\). An extended network is formed by adding exterior channels $\{Z_j\}$ to the initial causal setting, each generated in a similar fashion to $X_i$ but not designed to have cause-effect connections with the original pairs. These outer signals form a large repository of possible confounders. The final set contains $N_{\text{expanded}}$ time series, with the first $N$ holding the $N/2$ cause-effect relationships and the additional $N_{\text{external}}$ acting as external influencing signals that can inject disturbances into the original structure. 

\subsection{Linear Scheme}
\label{subsection:LinScheme}
In order to simulate the confounding effect of a high-dimensional network on the initial channels, linear influences are introduced from a subset of the additional channels \(\{Z_j\}\) to blur the association between \(X_i\) and \(Y_i\). We select a set of connected nodes $\mathcal{I} \subset \{1, \ldots, N_{ext}\}$. For each initial channel $Z_j(t)$, $j = 1, \ldots, N_{ext}$, we randomly select $k$ influencing nodes from $\mathcal{I}$ and assign influence weights $W_{ij}$ drawn from a uniform distribution. Then, the linear influence (i.e., \( L_i(t) = \sum_{j \in \mathcal{I}_i} W_{ij}Z_j(t) \)) on channel $i$ at time $t$ can be given as:
\[
{X}^*_i(t)
= X_i(t) + \sum_{j \in \mathcal{I}_i} W_{ij}Z_j(t),
\]
\[
{Y}^*_i(t)
= Y_i(t) + \sum_{j \in \mathcal{I}_i} W_{ij}Z_j(t),
\]

where $\mathcal{I}_i$ is the set of influencing nodes for channel $i$. For each initial channel $X^*_i(t)$ and each influencing node \(Z_j(t) \in \mathcal{I}_i\), we assign an influence weight \(W_{ij}\) drawn from a uniform distribution centered around a specified influence weight \(\omega\) ( i.e., \(W_{ij} \sim \text{Uniform} (\omega - \frac{\omega}{16},  \omega + \frac{\omega}{16}) \)). The modified signals \({X}^*_i\) and \({Y}^*_i\) retain their causal arrangement in principle, but it is obscured by superimposed signals from the outer set. In the case of large amplitude weights (i.e., higher connectivity), the underlying causal association will be difficult to detect unless these background effects are adequately addressed.

\subsection{Nonlinear Scheme}
\label{subsection:NonLinScheme}
In the nonlinear scheme, we introduce moderate nonlinearity into the initial channels. The outer channels \(\{Z_j\}\) remain potential confounders, but each channel \({X}^*_i\) or \({Y}^*_i\) includes products \(Z_j(t) Z_k(t)\) and a smooth function of linear influence \( L_i(t) = \sum_{j \in \mathcal{I}_i} W_{ij}Z_j(t)\).  The nonlinear influence on channel $i$ at time $t$ can be exemplified as:
\[
{X}^*_i(t) = X_i(t) + f\Bigl(\sum_{j \in \mathcal{I}_i}W_{ij} Z_j(t)\Bigr) +  \omega^* \sum_{j \in \mathcal{I}_i}\sum_{k>j}W_{ij} W_{ik} Z_j(t) Z_k(t),
\]

\[
{Y}^*_i(t) = Y_i(t) + f\Bigl(\sum_{j \in \mathcal{I}_i}W_{ij} Z_j(t)\Bigr) +  \omega^* \sum_{j \in \mathcal{I}_i}\sum_{k>j}W_{ij} W_{ik} Z_j(t) Z_k(t),
\]

where \(f(x)=\ln\bigl(1+e^x\bigr)\) and \(\omega^*\) scales the interaction term.

\subsection{Causal Scheme}
\label{subsection:CausalScheme}
A separate simulation setup imposes cause-effect links among initial nodes and uses a broader set of \(\{Z_j\}\) to create a complex background. The outer channels modify each \(X_i\) or \(Y_i\) according to difference equations like
\[
{X}^*_i(t) = \mu X_i(t-1) + \sum_{j\in \mathcal{I}_i}W_{ij} Z_j(t-1) +  \nu_i(t), \text{ where } \nu_i(t) \sim \mathrm{WN}(0,\sigma^2_\nu)
\]
\[
{Y}^*_i(t) = \mu Y_i(t-1) + \sum_{j\in \mathcal{I}_i}W_{ij} Z_j(t-1) +  \nu_i(t), \text{ where } \nu_i(t) \sim \mathrm{WN}(0,\sigma^2_\nu)
\]
making it difficult to recover the direct link \(X_i \to Y_i\) unless one accounts for external influences. This form is used as a framework in which multiple strategies can be compared.

\subsection{Simulation Results}

\subsection*{Linear and Nonlinear Schemes of Influences}

\begin{figure}[H]
    \centering
    \includegraphics[trim= 0 0 8cm 0,clip,width=0.48\linewidth]{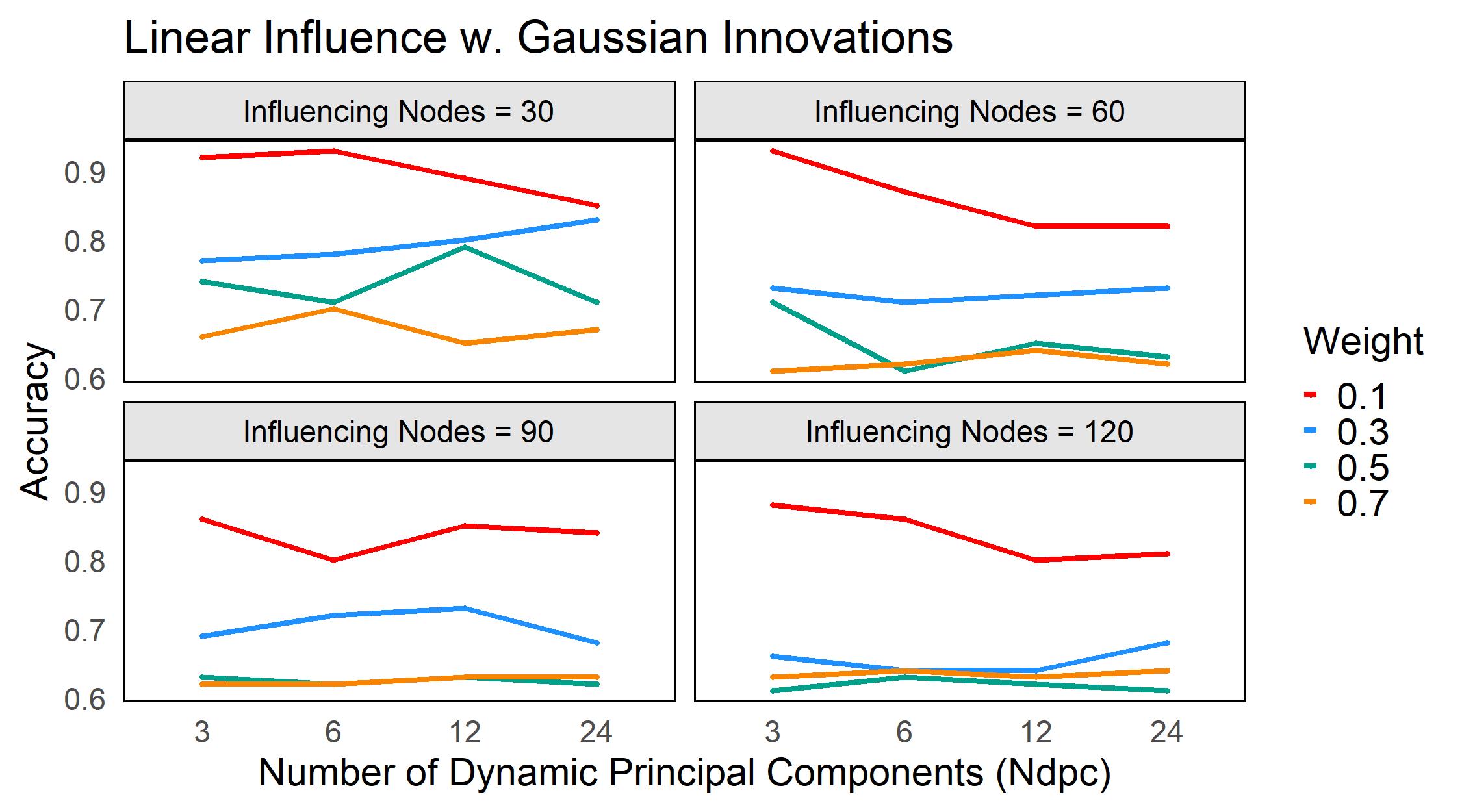}
    \includegraphics[trim= 5cm 0 0 0,clip,width=0.5075\linewidth]{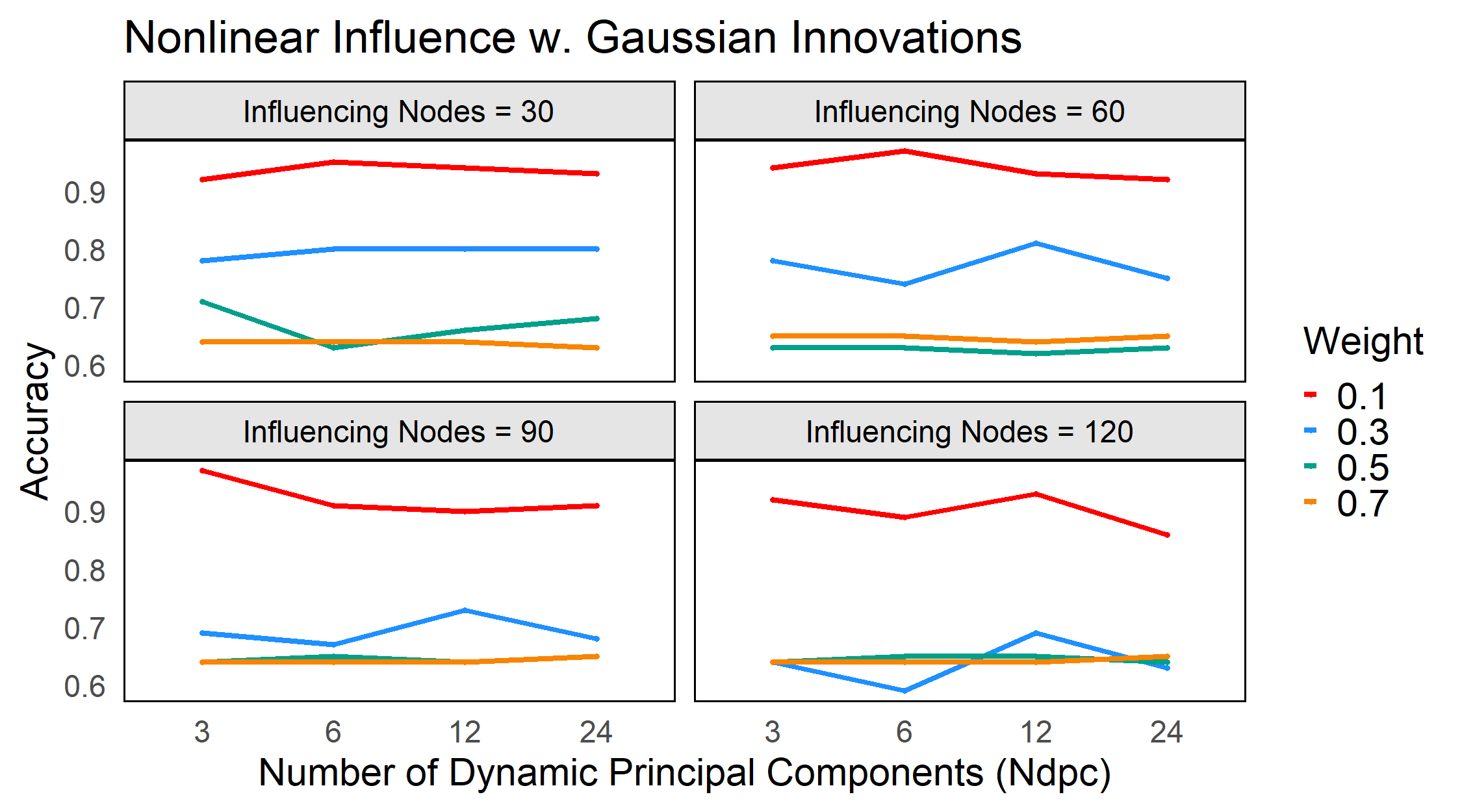}\\[1em]

    \includegraphics[trim= 0 0 8cm 0,clip,width=0.48\linewidth]{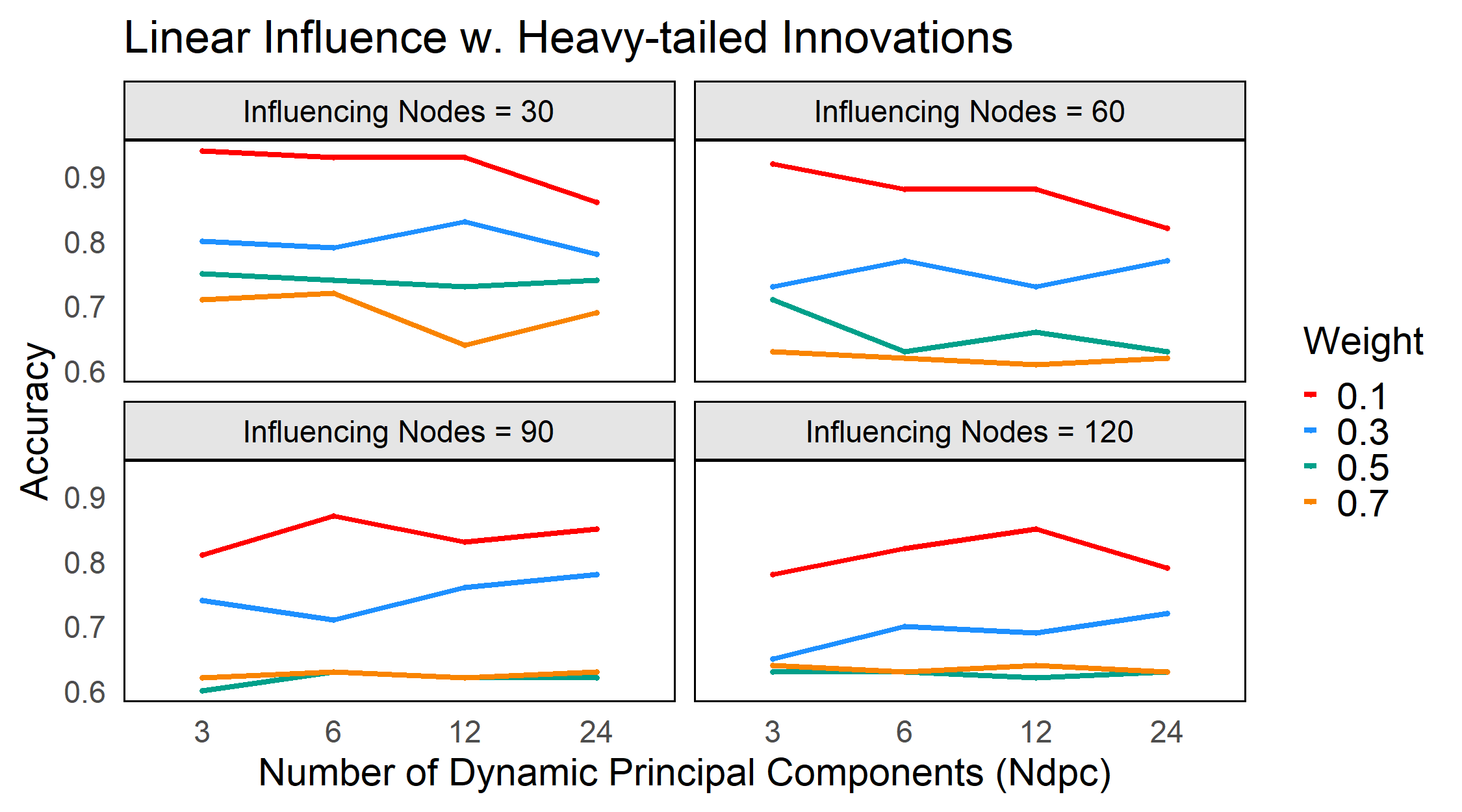}
    \includegraphics[trim= 5cm 0 0 0,clip,width=0.5075\linewidth]{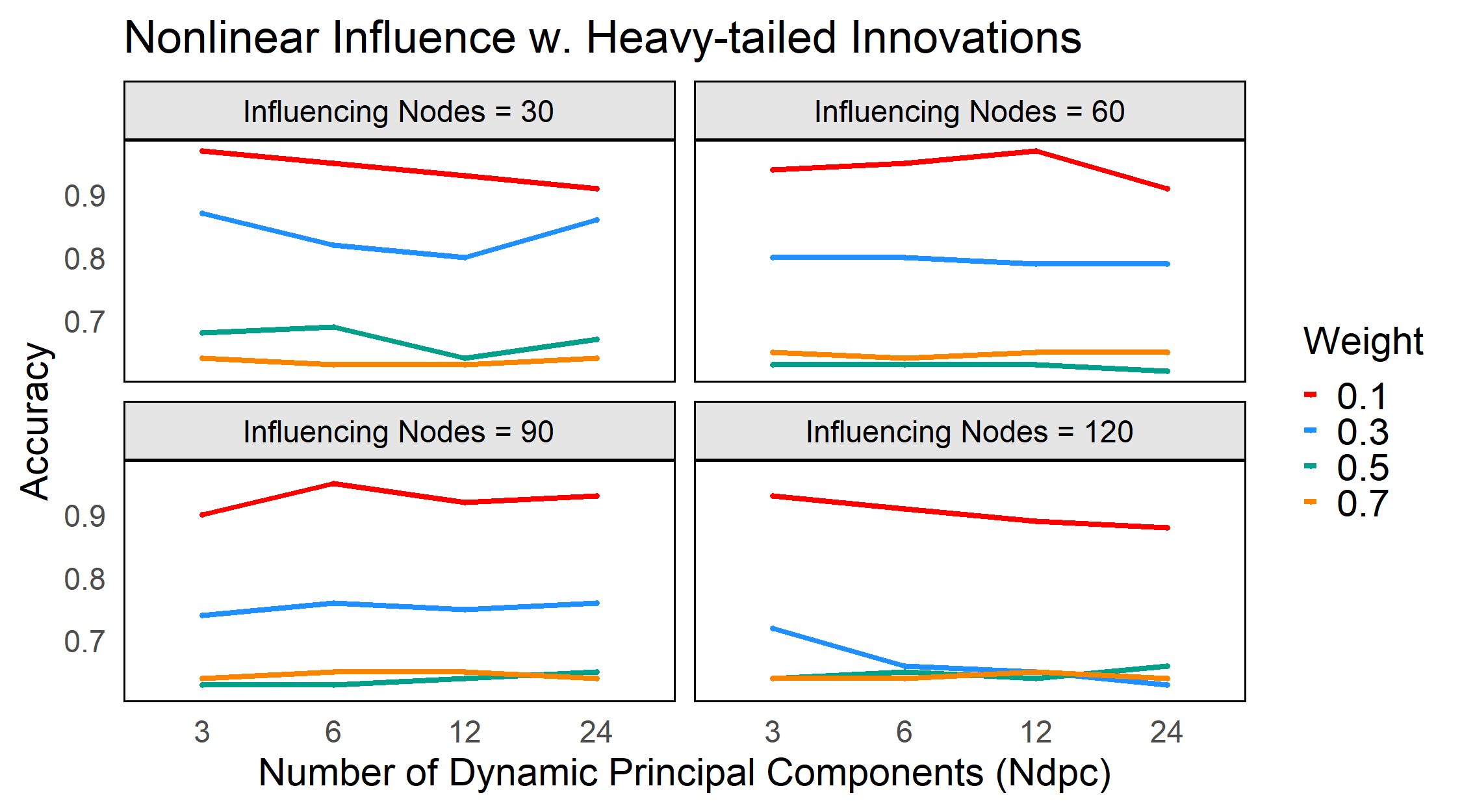}\\

    \caption{Accuracy results under Linear and Nonlinear Influences}
    \label{fig:Sims}
\end{figure}

We assessed the proposed approach’s ability to recover known causal connections in simulated high-dimensional networks having both linear and nonlinear interference from outer channels explained in \ref{subsection:LinScheme} and \ref{subsection:NonLinScheme}. In these settings, different numbers of influencing nodes (with varying weights) on the COI in a high-dimensional network are introduced. The size of the networks is set to 512 nodes. Thus, the COI consists of the first 40 (i.e., $N = 40$) nodes out of a total of 512 (i.e., $ N_{expanded} = 512$) nodes. Figure \ref{fig:Sims} summarizes the results across these multiple setups. Results show that the procedure generally maintains high accuracy when confounding is modest. As the influence weight and network density grow, the method’s performance declines and stabilizes, indicating that strong external perturbations and extensive connectivity cause significant challenges. Increasing the number of dynamic principal components beyond a small set does not substantially improve accuracy under heavier confounding. However, the number of scores can be chosen according to their ability for the total variance explainability. In these scenarios, we focused more on a relatively small number of scores rather than explainability, so, total variance explainability changes between $20\%$ to $65\%$. Overall, the results suggest that spectral dynamic principal component analysis can effectively isolate the channels of interest in large networks, though very high connectivity or influence weights reduce its success.

The method achieved high accuracy (above 0.90) when influence weights were low (e.g., 0.1), even as the number of interfering nodes increased from 30 to 120. As the influence weight rose to 0.3, performance remained acceptable for smaller interfering sets but declined with increasing network density.

When the influence weights reached 0.5 and 0.7, the accuracy obtained was around 0.64–0.65, unaffected by the number of dynamic principal components used. This suggests that under strong nonlinear disturbances and widespread or super connectivity, the method’s ability to disentangle the channels of interest becomes limited.

In lower interference regimes, sDPCA consistently filtered out background effects and preserved meaningful causal relationships. However, in highly entangled networks with amplified nonlinear effects, performance deteriorates regardless of the dimensionality reduction depth. These findings point to the method’s strengths in moderately complex environments and its limitations in ultra-dense or strongly nonlinear settings.

\subsection*{Causative Schemes of Influences}

\begin{figure}[htbp]
    \centering
    \includegraphics[width=1\linewidth]{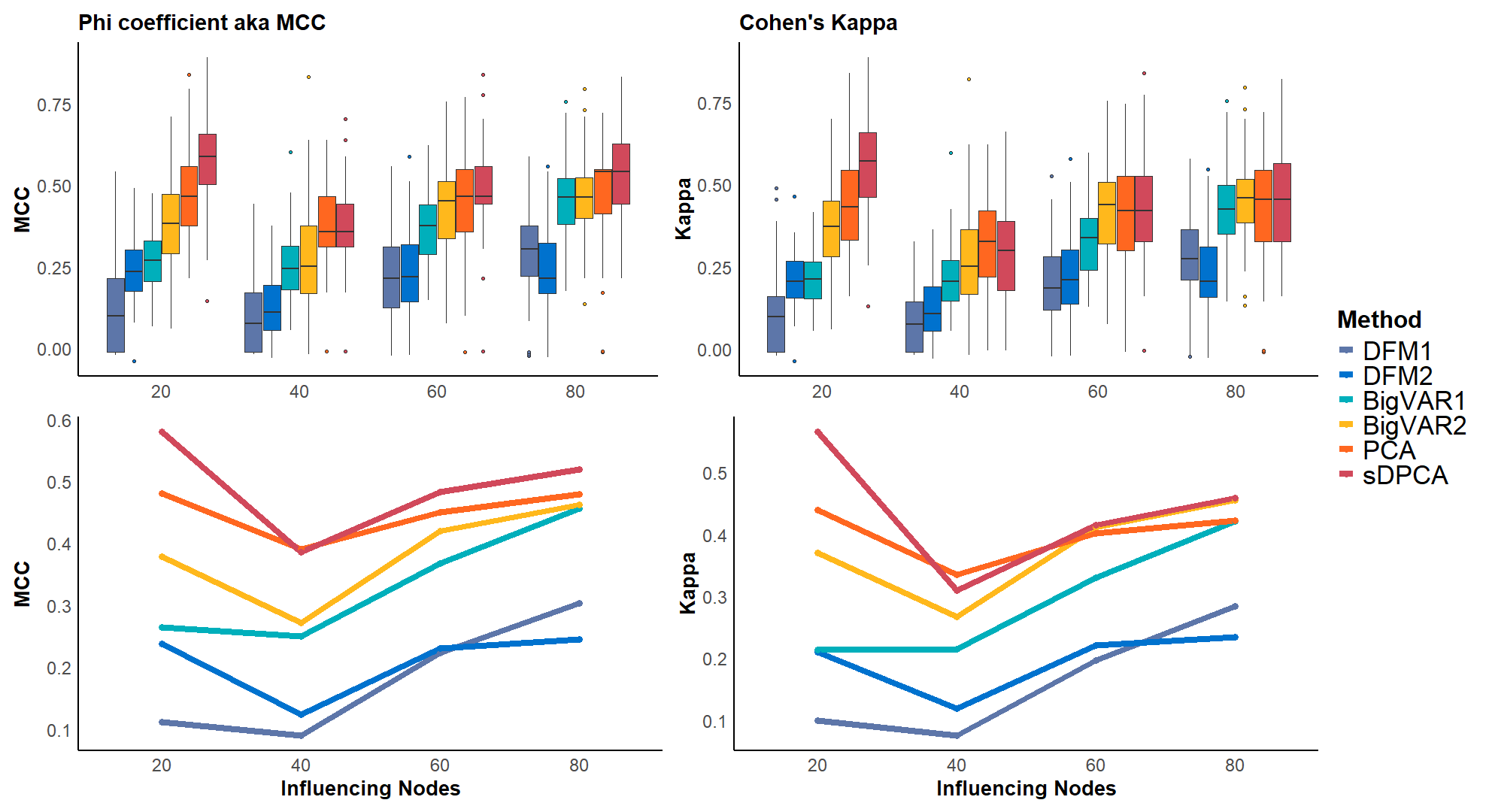}
    \caption{Comparison of accuracies under Causative Influences}
    \label{fig:Comparison}
\end{figure}

In another simulation scenario, we consider the case where the outer network acts in a causative manner given in \ref{subsection:CausalScheme}. The size of the networks is set to 100 nodes in this case. The COI consists of the first 20 (i.e., $N = 20$) nodes out of a total of 100 (i.e., $ N_{expanded} = 100$) nodes. Figure \ref{fig:Comparison} shows boxplots of the Matthews correlation coefficient (MCC) and Cohen’s Kappa in the upper panels, along with line plots of mean values in the lower panels. Each plot illustrates how well different methods detect the injected causal links under varying degrees of network density (low, moderate, and high connectivity). The horizontal axis indicates the number of influencing nodes, which grows from low to high levels of confounding. GC–PCA uses standard principal component analysis to remove shared background signals before testing for Granger causality. DFM1, the default procedure, extracts idiosyncratic components via dynamic factor modeling and fits a sparse VAR to interpret nonzero coefficients as causal links. DFM2 modifies this by applying a formal Granger test to the idiosyncratic signals, rather than relying solely on a coefficient matrix. BigVAR1 fits an elastic-net penalized VAR to the influenced network, while BigVAR2 uses the same penalization strategy but only on factor-based idiosyncratic components extracted, forming a hybrid of DFM and penalized VAR estimation.

Granger causality combined with sDPCA tends to provide higher accuracy scores across all scenarios. GC-PCA also performs well, though becomes less noticeable when the network density increases. Unlike the previous scenario, here, the number of scores used in partialling out of the effect of the outer network was determined when the cumulative explained variance reached $75\%$. Fitting BigVAR to only the idiosyncratic components estimated via DFM (BigVAR2) improves performance relative to fitting it to the full network (BigVAR1). DFM1 (default mode) shows relatively low performance in these simulations, whereas DFM2 (a conventional test for GC on the idiosyncratic components) recovers more causal links but remains below the top-performing procedures. Accuracy measures decrease as the number of influencing nodes grows, reflecting the increased difficulty of recovering true causal connections in super-connected systems. Methods that factor out or model the common background, such as sDPCA or the hybrid approach in BigVAR2, tend to be more robust in this regard.


\section{Application to EEG data }

In this section, we apply our proposed approach to EEG recordings from an experiment involving motor execution and imagery tasks, offering insights into the temporal dynamics of brain activity. EEG signals were collected from multiple subjects using the BCI2000 system with a 64-channel setup; the data is publicly available at PhysioNet.org \citep{schalk2004bci2000}. Each subject completed 12 task-oriented runs. Participants either physically performed or imagined movements in response to visual cues on a screen, such as opening and closing fists or feet, depending on the target's position. The tasks were divided into four categories: physical movements of the left or right fist, imaginary movements of the left or right fist, physical movements of both fists or both feet, and imaginary movements of both fists or both feet. EEG data were sampled at 160 Hz, along with an annotation channel indicating rest or the onset of motion (real or imagined) for specific movements. The EEG montage followed the international 10-10 system, omitting certain electrodes (Nz, F9, F10, FT9, FT10, A1, A2, TP9, TP10, P9, and P10).

All data underwent pre-processing to filter out artifacts like eye blinks and muscle movements, preserving the temporal structure of the recordings. The procedure included a high-pass filter at 0.5 Hz to eliminate low-frequency noise, removal of 60 Hz AC line noise, and Independent Component Analysis (ICA) using MNE-Python routines \citep{GramfortEtAl2013a, larson_2024_10519948} to isolate and exclude artifact components. 

\begin{figure}[ht!]
\centering

\begin{subfigure}[b]{0.325\linewidth}
  \includegraphics[width=\linewidth,
                   trim=170pt 125pt 140pt 60pt,
                   clip]{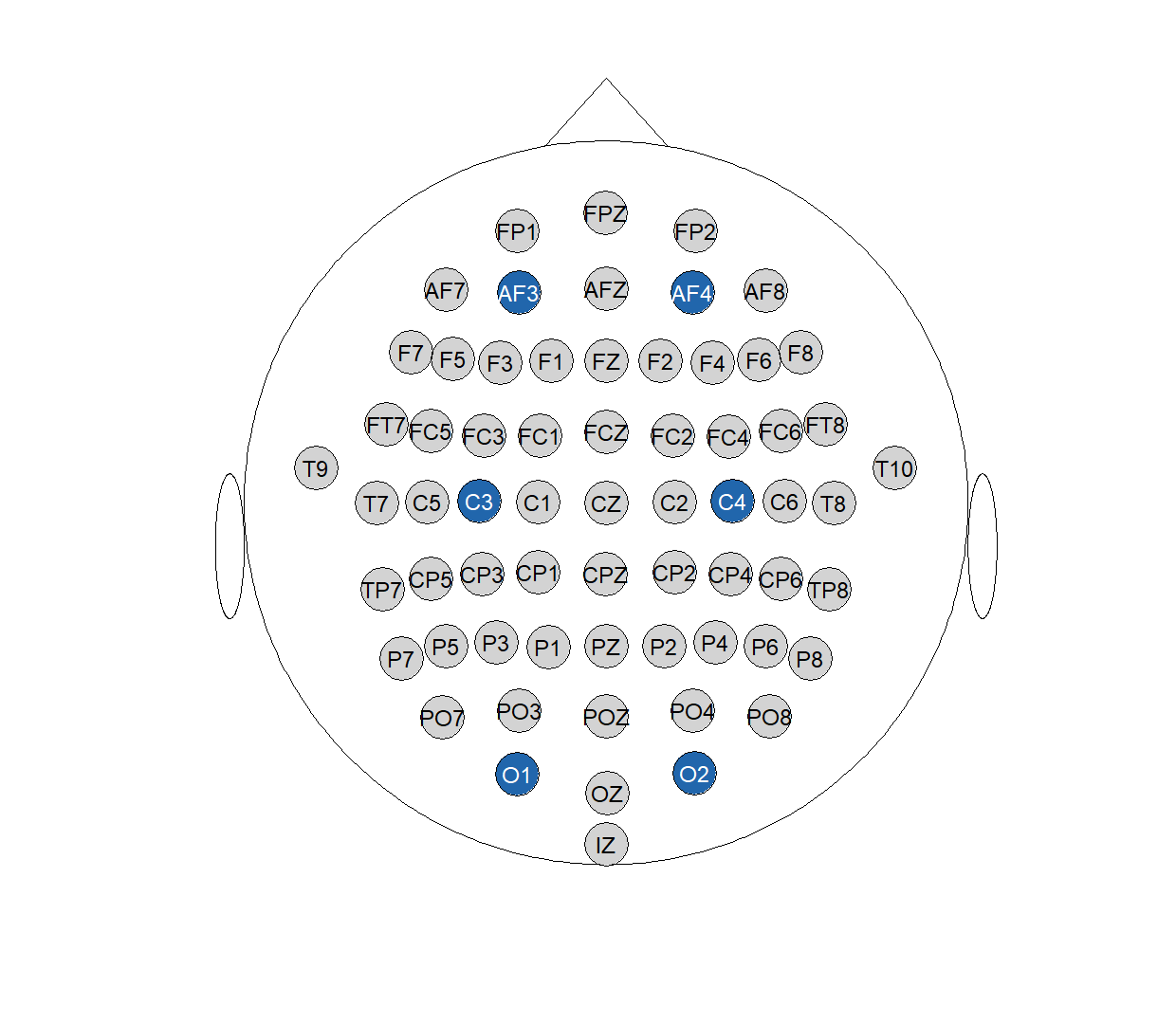}
  \caption{}
  \label{fig:COI_a}
\end{subfigure}
\quad
\begin{subfigure}[b]{0.60\linewidth}
  \includegraphics[width=\linewidth]{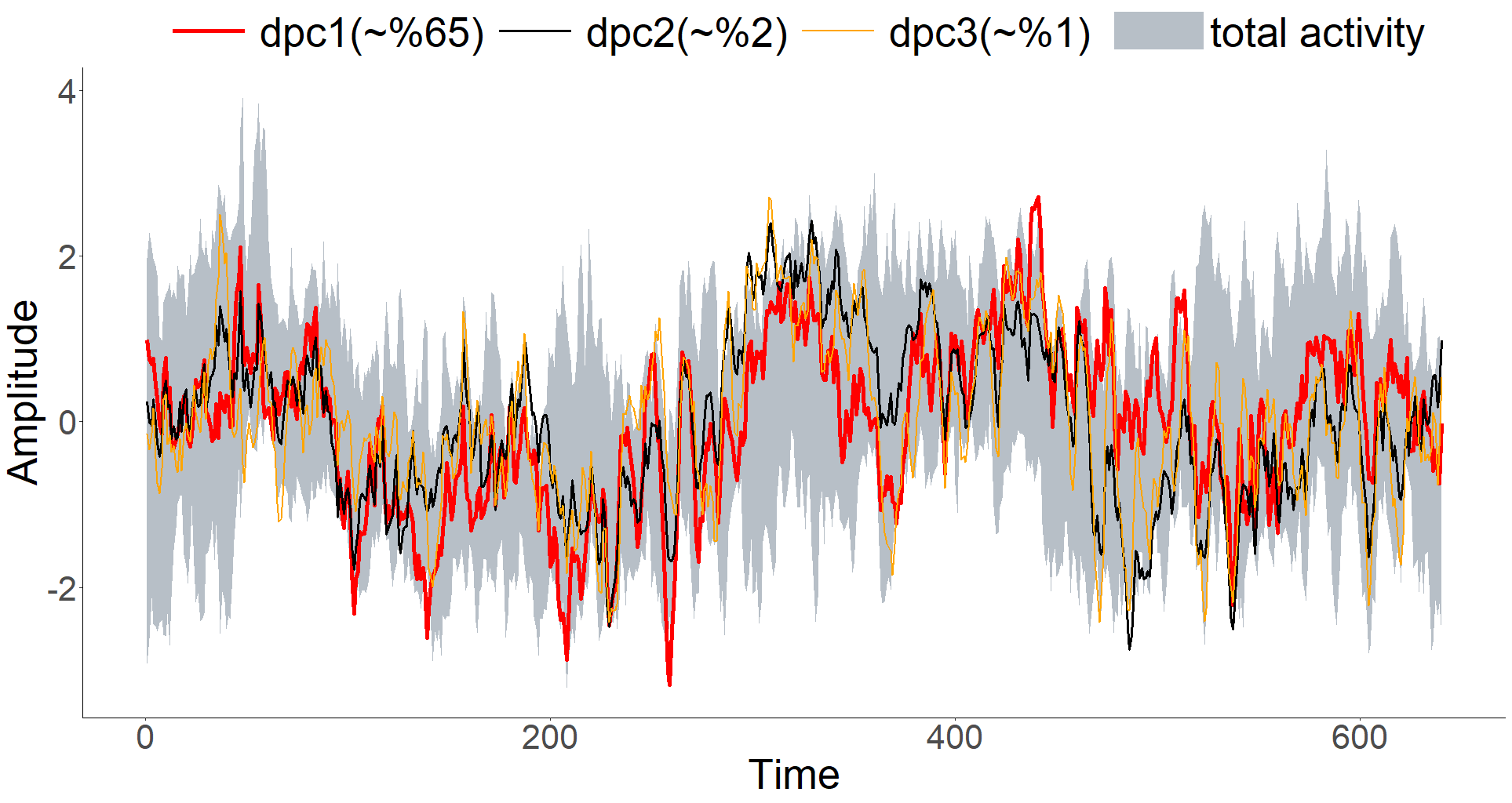}
  \caption{}
  \label{fig:COI_b}
\end{subfigure}

\caption{(a)~\bluesquare: COI, \lightgraysquare : Excluded nodes  (b)~sDPCA scores from Excluded nodes}
\label{fig:COI}
\end{figure}

Despite standard pre-processing, some recordings had unresolved issues like sawtooth traces, excessive blinking, broken channels, and unbalanced epochs. Therefore, we limited our analysis to 35 right-handed subjects (17 females and 18 males) whose recordings were qualified. To demonstrate how the proposed method can dissect causal connections between pairwise nodes in a network, we focused on six channels of interest (COI) out of 64 EEG channels: AF3 and AF4 positioned over the pre-frontal cortex, C3 and C4 placed over the motor cortex, and O1 and O2 over the occipital cortex, as illustrated in Figure \ref{fig:COI_a}. These channels were chosen because they correspond to brain regions likely to be active during voluntary movement, imagery, and visual perception.

On the other hand, another important objective is to monitor the effect of the dimension reduction utilizing the sDPCA and its impact on the isolated channels. In particular, we propose that basic statistical tools such as trace plots, serial and partial autocorrelation functions, spectrum estimates, and many more can be practically used to determine the extent to which the method alters time-dependent properties in time series. This helps us to assess the efficacy of the proposed methodology for isolating COI while maintaining the temporal characteristics of the signals.

In Figure \ref{fig:COI_b}, we illustrate an example of one epoch from the dataset. Three sDPCA scores obtained from excluded nodes are plotted over the time series data from which they were derived. This illustrates that the small number of scores can capture the global modes and preserve temporal structures/dynamics of the time series.

\begin{figure}[H]
\begin{tabular}{>{\centering\arraybackslash}m{3.15cm}>{\centering\arraybackslash}m{3.15cm}>{\centering\arraybackslash}m{3.15cm}>{\centering\arraybackslash}m{3.15cm}}
 \makecell{\includegraphics[width=1\linewidth]{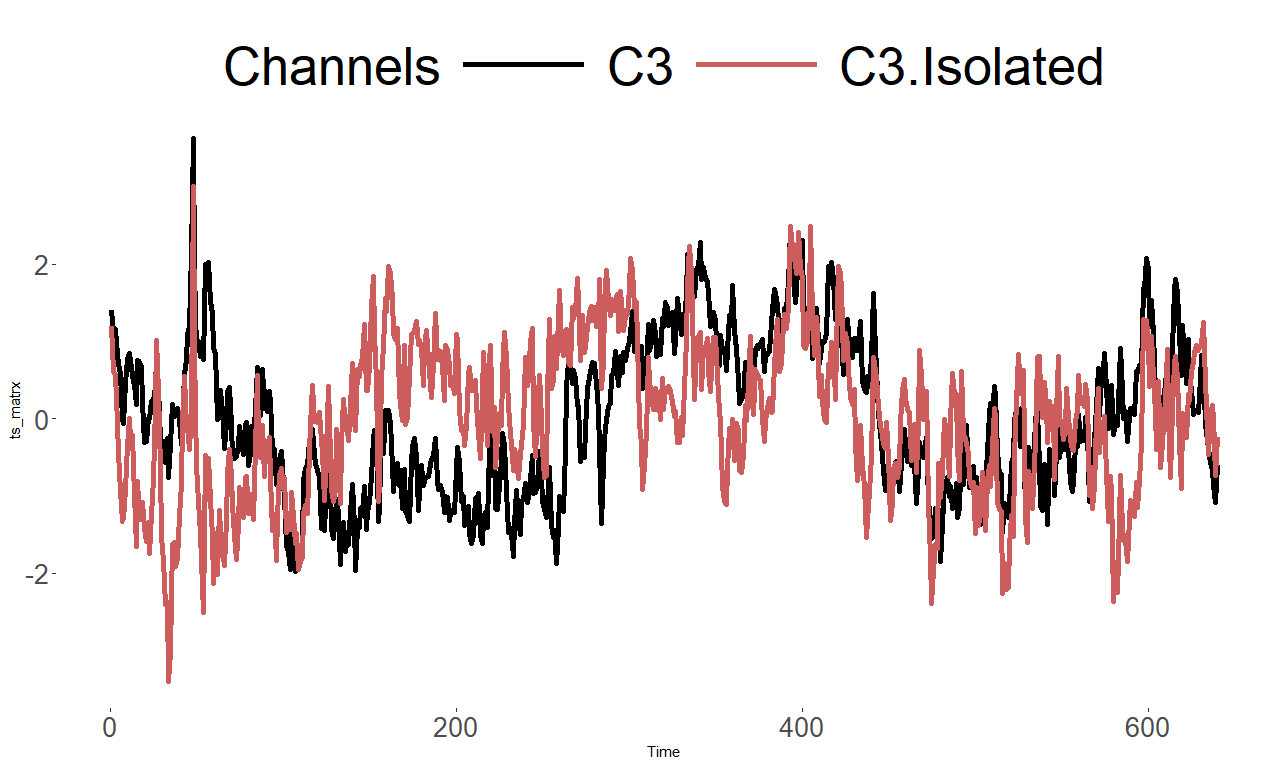} \\ \includegraphics[width=1\linewidth]{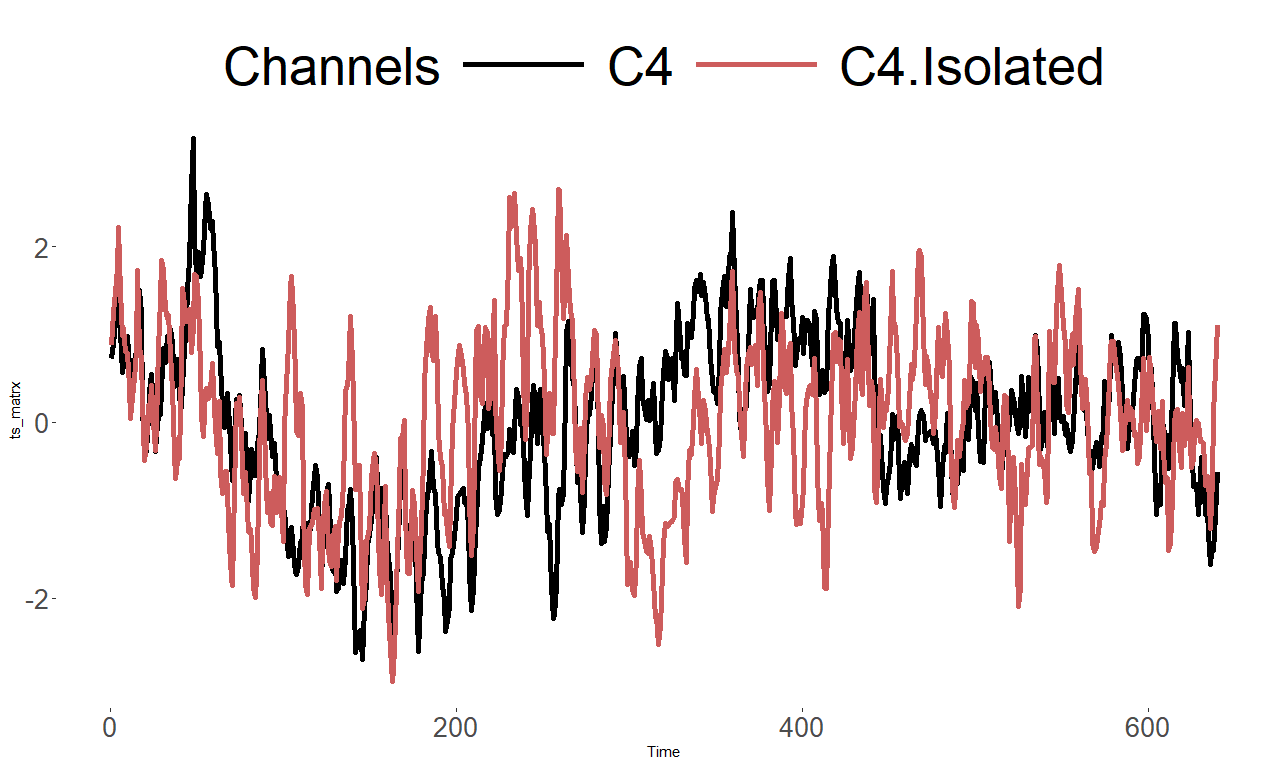}}
 & \makecell{\includegraphics[width=1\linewidth]{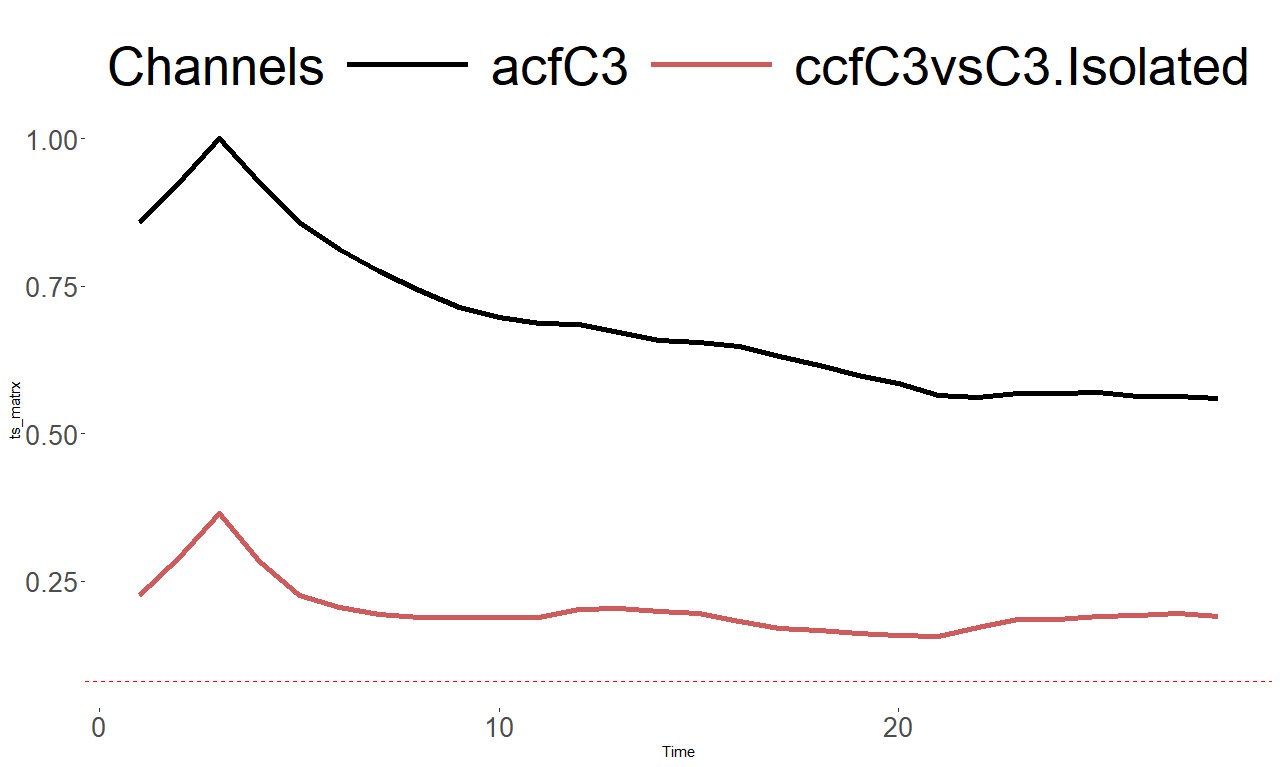} \\ \includegraphics[width=1\linewidth]{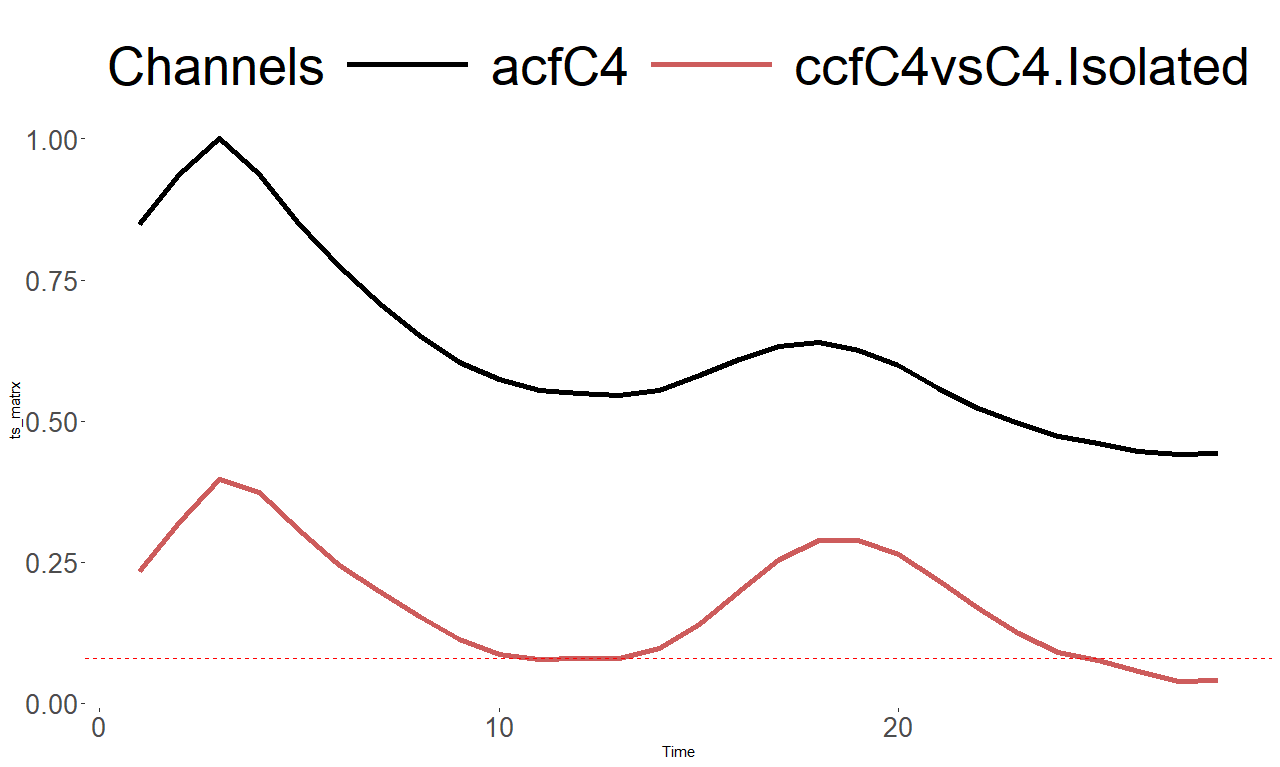}} & \makecell{\includegraphics[width=1\linewidth]{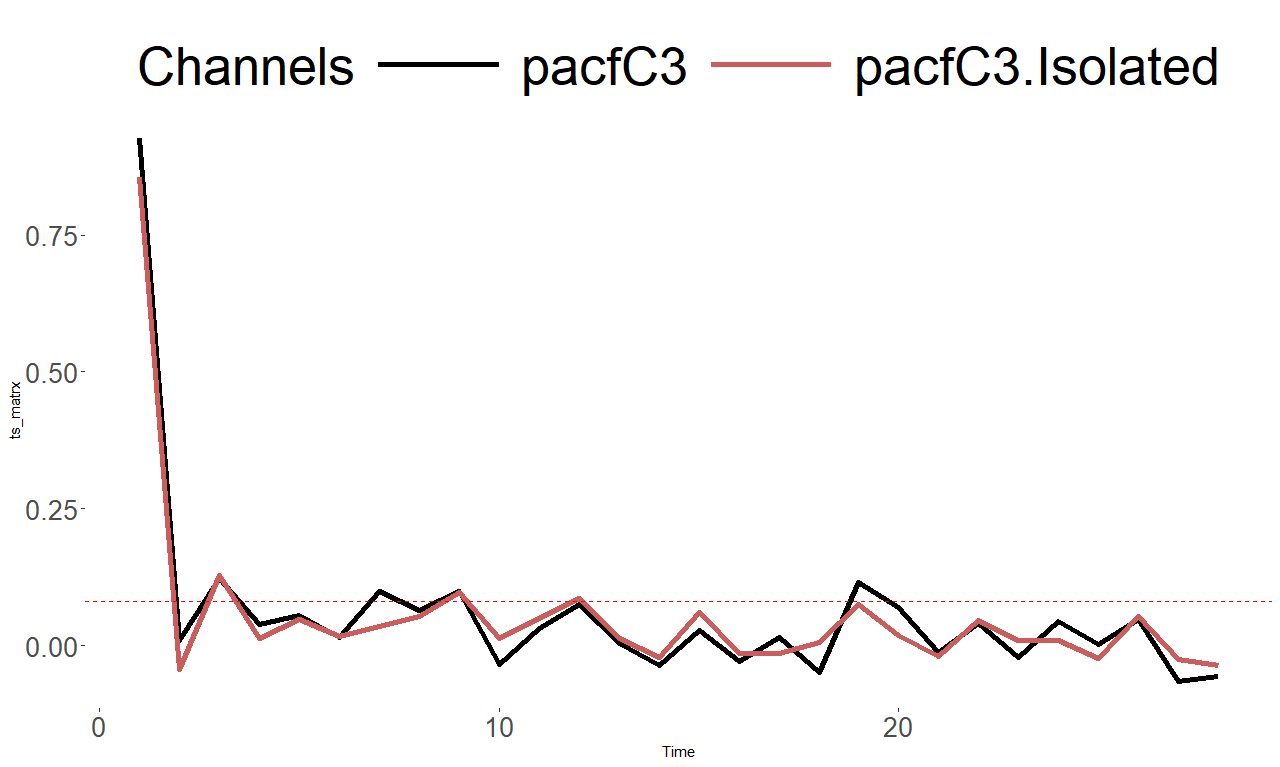} \\ \includegraphics[width=1\linewidth]{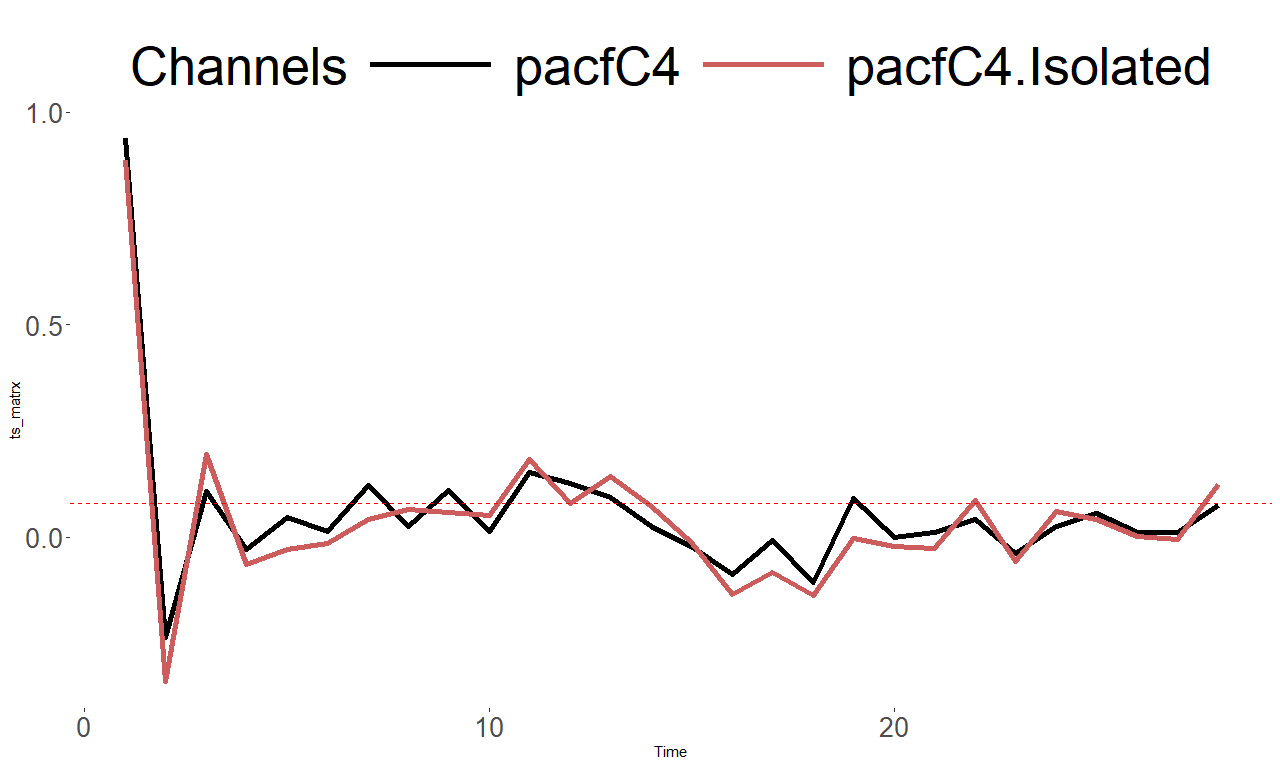}} & \makecell{\includegraphics[width=1\linewidth]{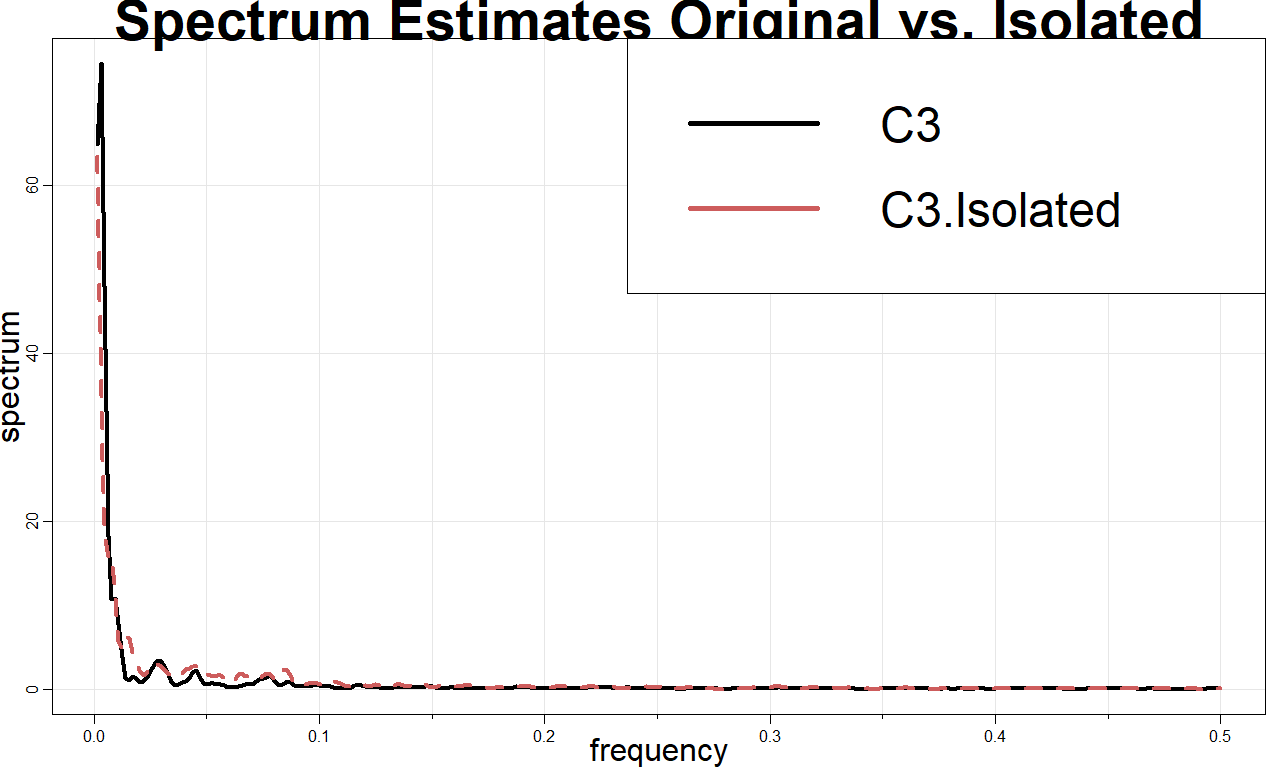} \\ \includegraphics[width=1\linewidth]{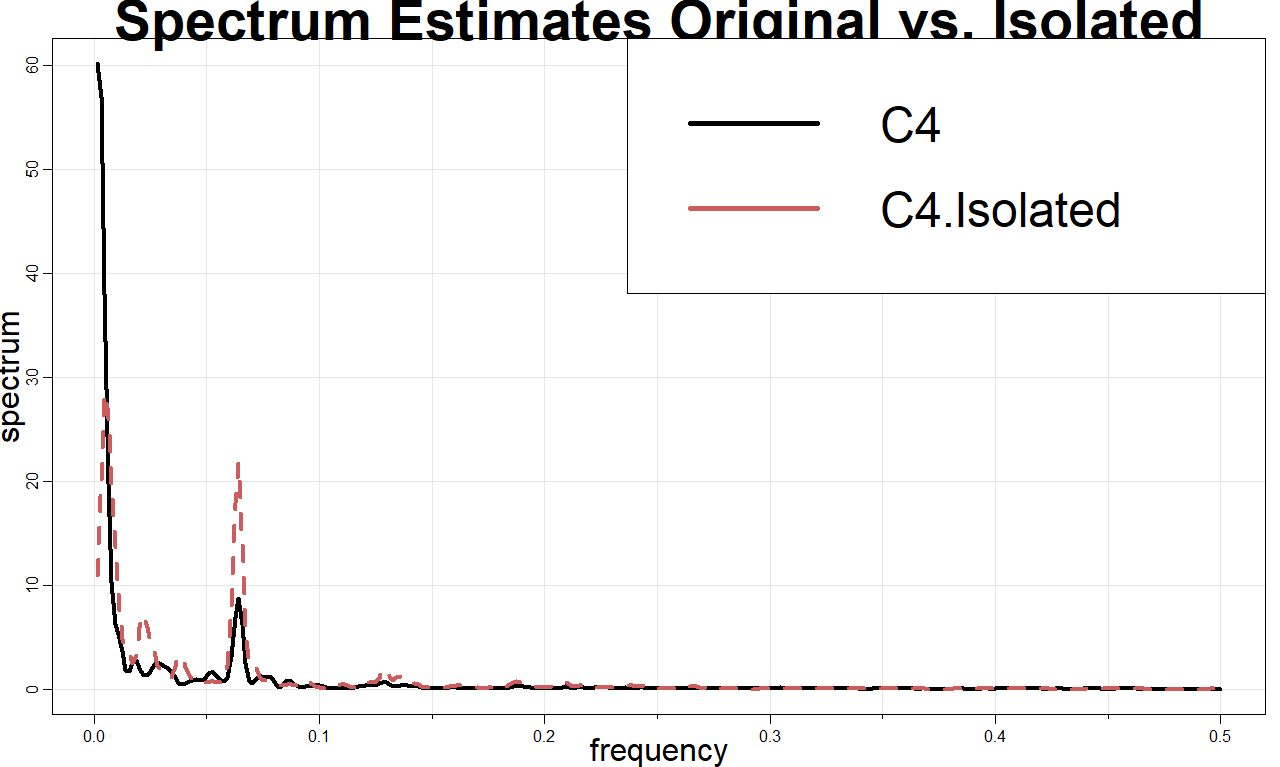}} \\
\footnotesize (a) trace & \footnotesize (b) acf & \footnotesize (c) pacf & \footnotesize (d) spectrum \\
\end{tabular}
  \caption{Sample vs Isolated channels C3 and C4.}
  \label{fig:C3vsC4}
\end{figure}

\begin{figure}[H]
\begin{tabular}{>{\centering\arraybackslash}m{3.15cm}>{\centering\arraybackslash}m{3.15cm}>{\centering\arraybackslash}m{3.15cm}>{\centering\arraybackslash}m{3.15cm}}
 \makecell{\includegraphics[width=1\linewidth]{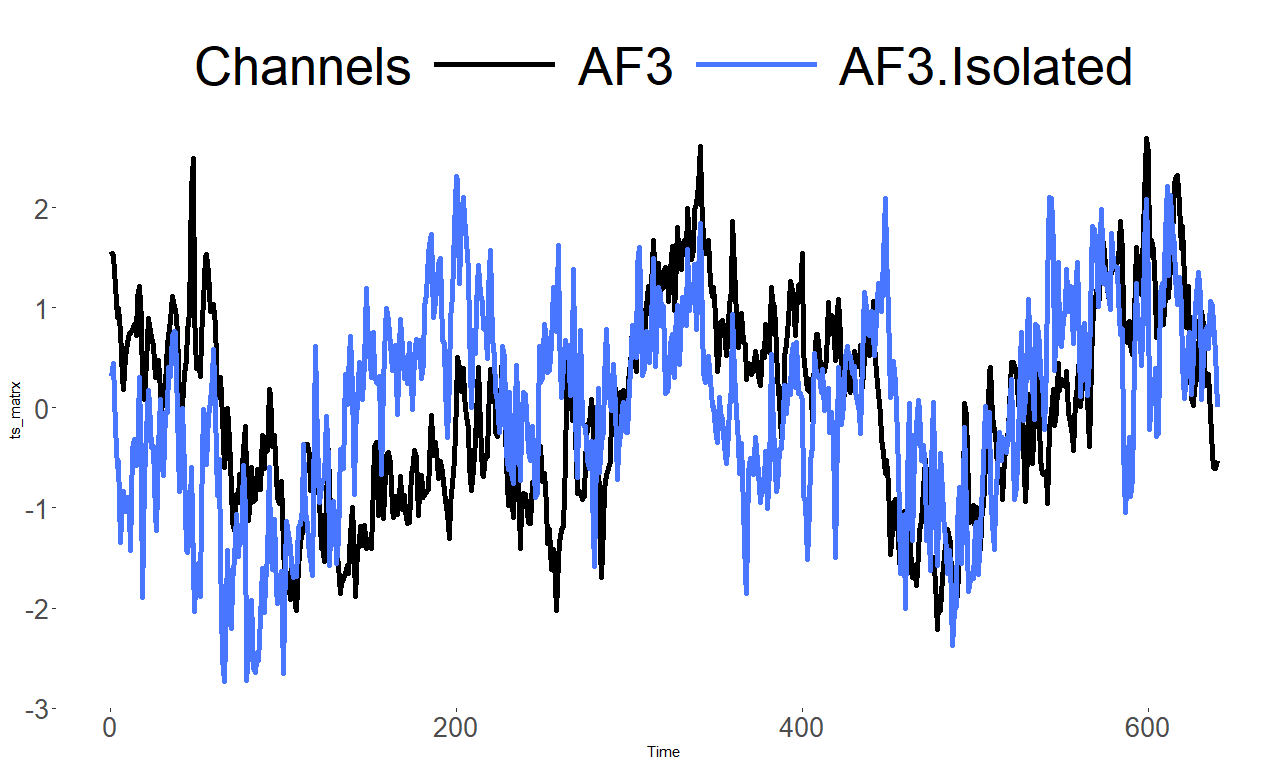} \\ \includegraphics[width=1\linewidth]{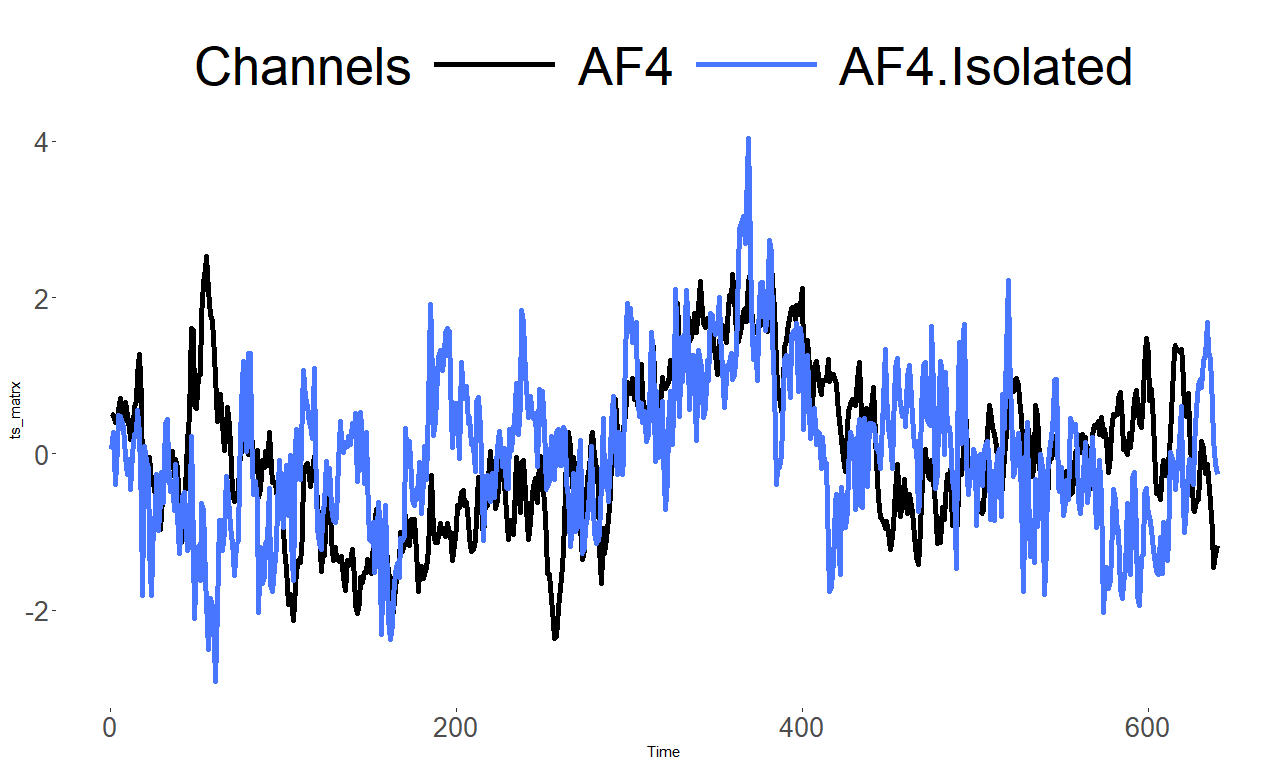}}
 & \makecell{\includegraphics[width=1\linewidth]{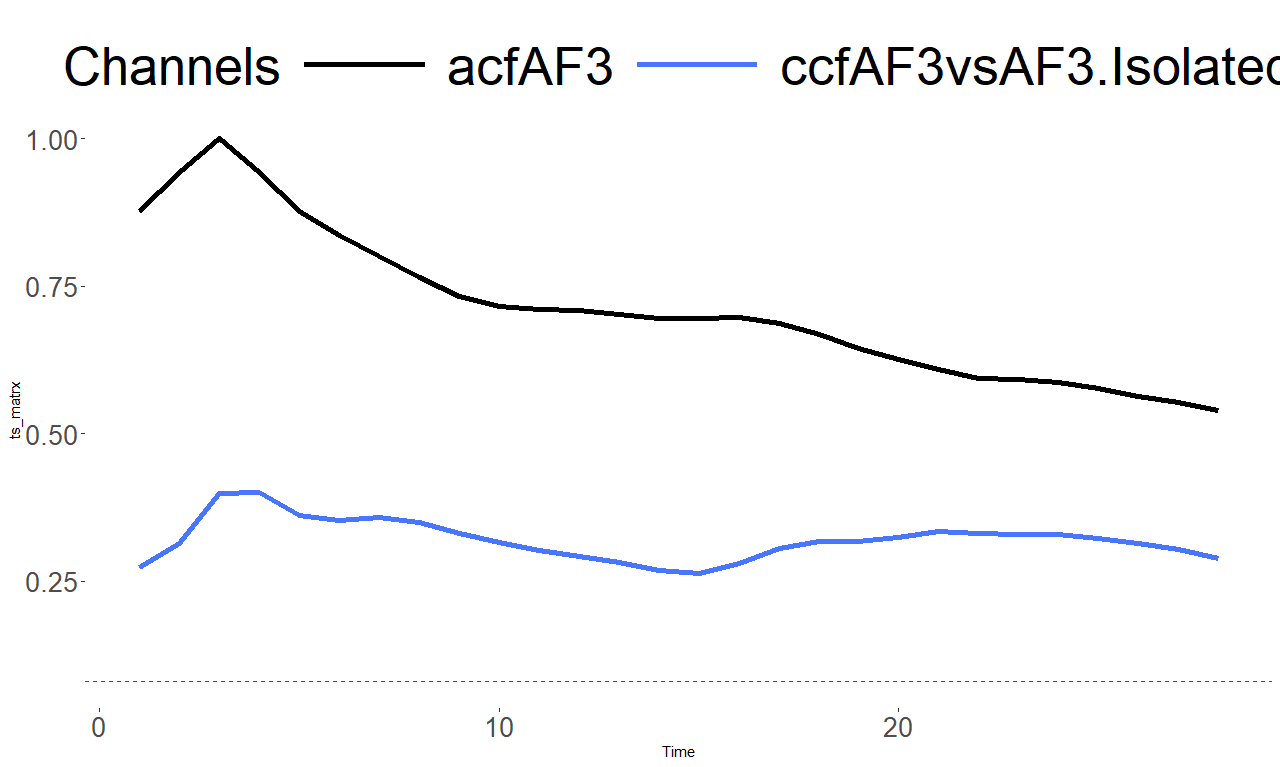} \\ \includegraphics[width=1\linewidth]{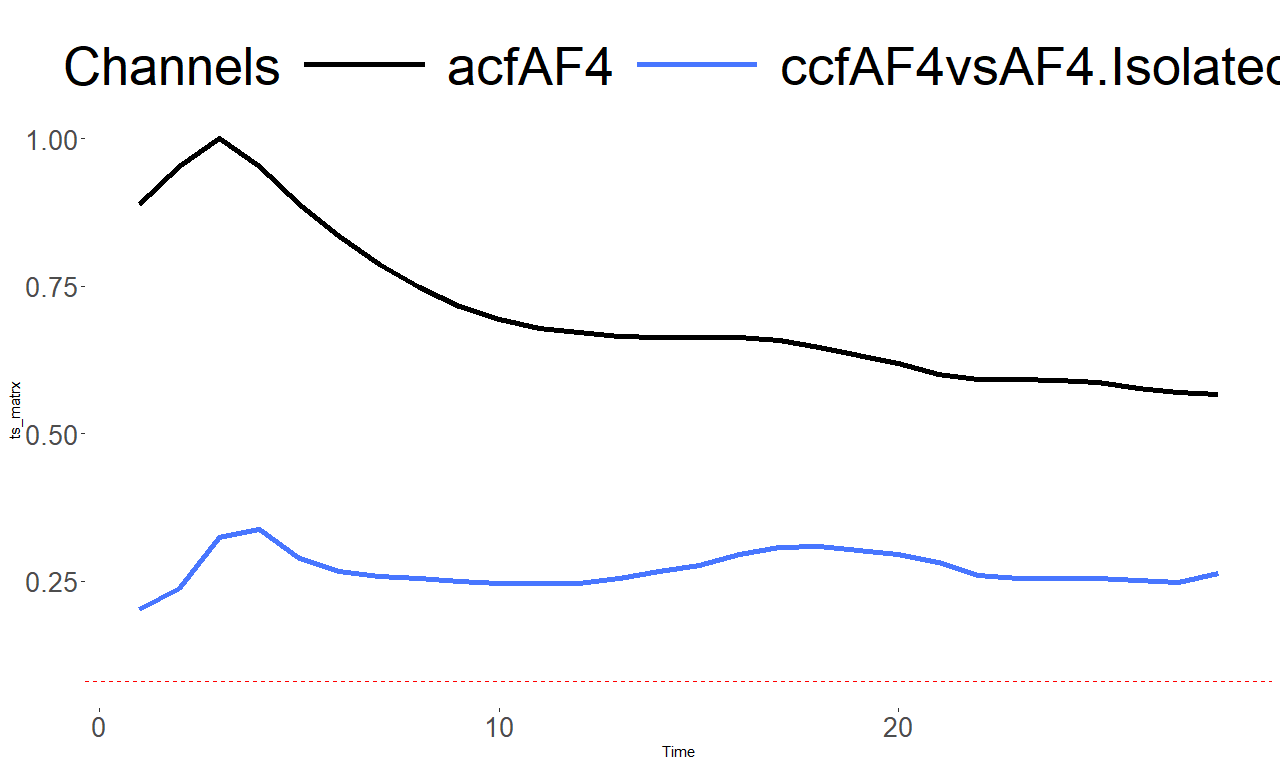}} & \makecell{\includegraphics[width=1\linewidth]{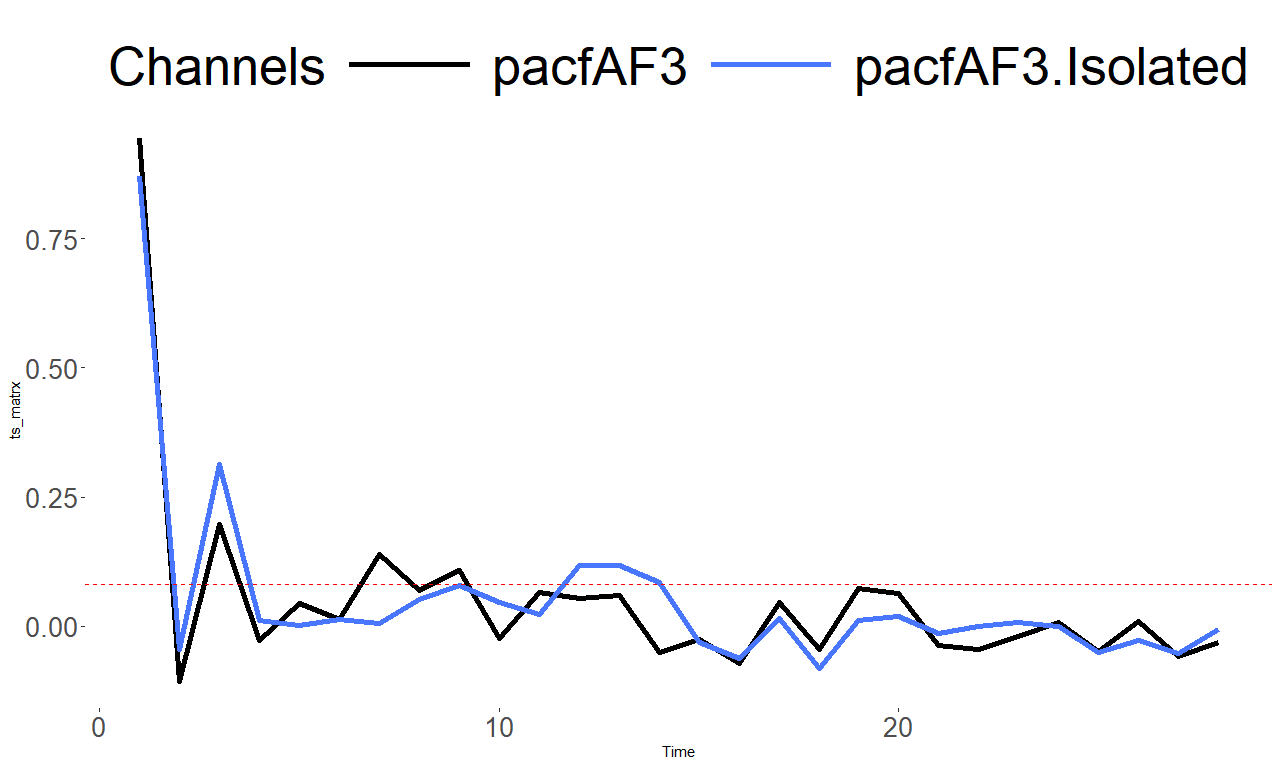} \\ \includegraphics[width=1\linewidth]{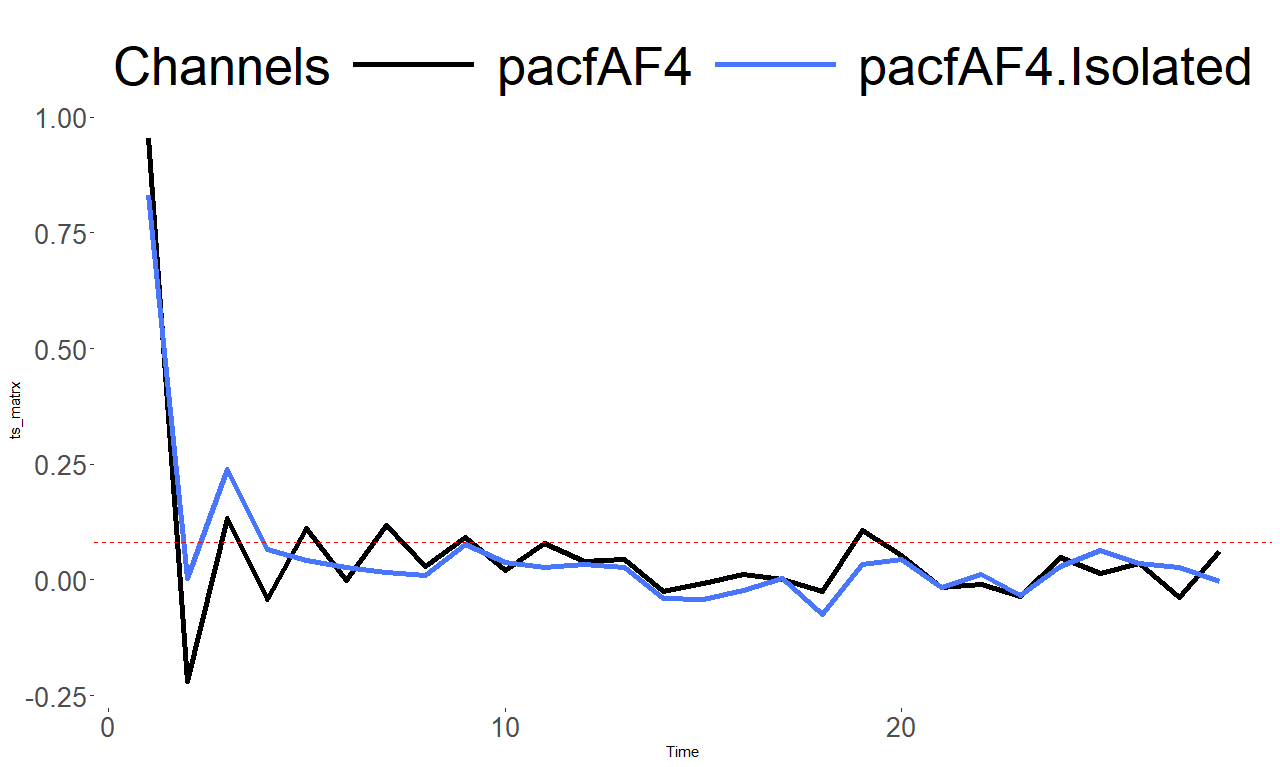}} & \makecell{\includegraphics[width=1\linewidth]{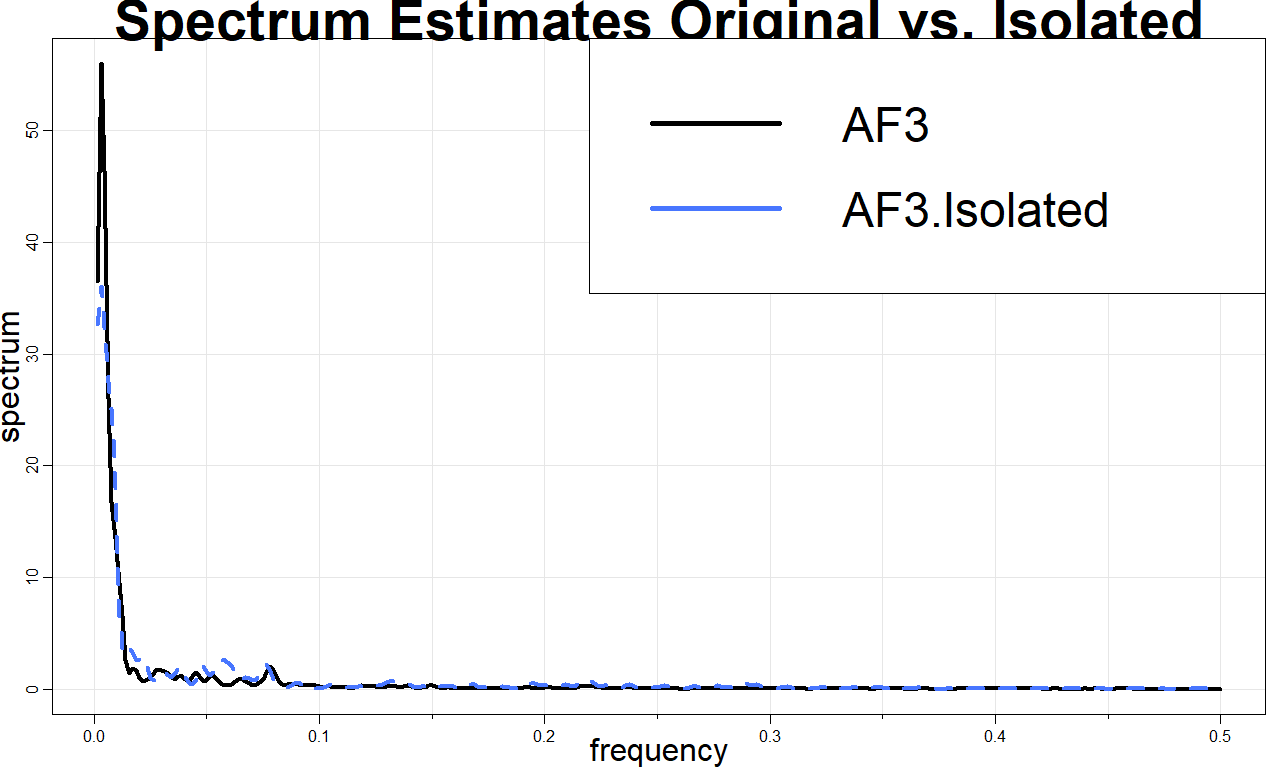} \\ \includegraphics[width=1\linewidth]{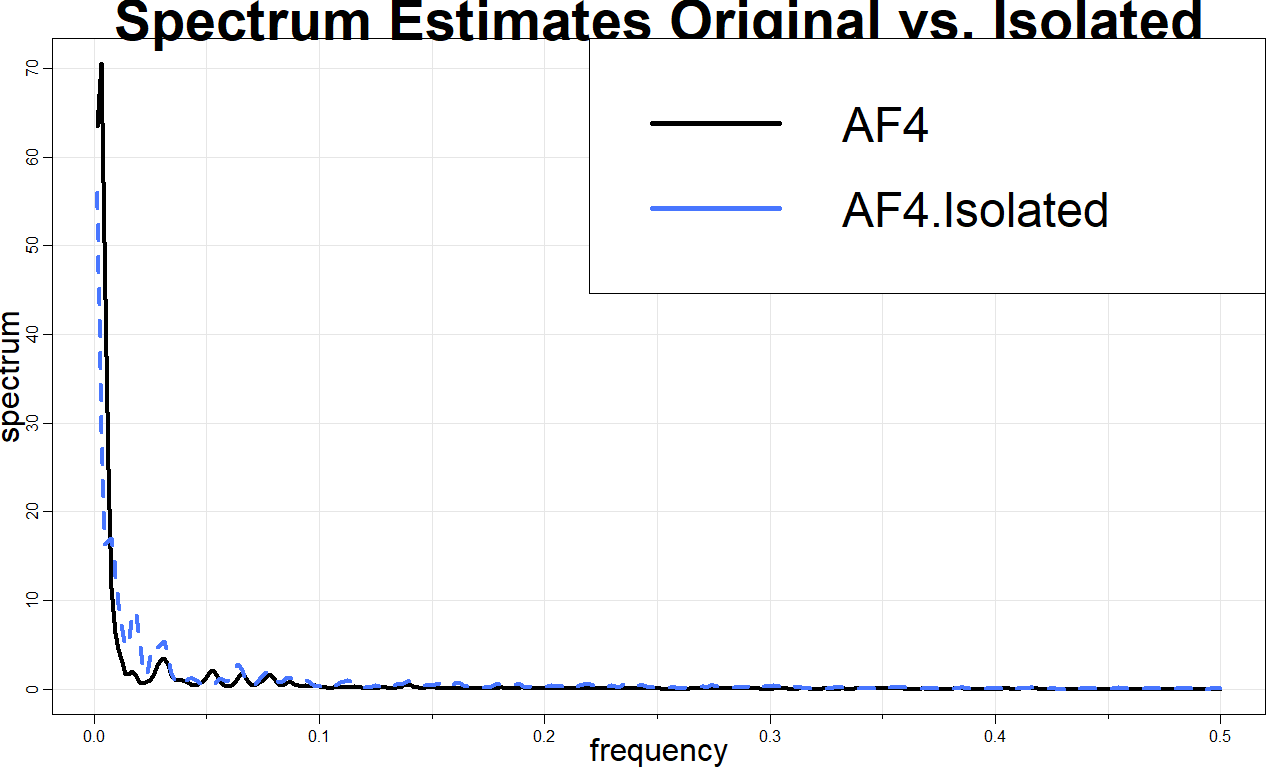}} \\
\footnotesize (a) trace & \footnotesize (b) acf & \footnotesize (c) pacf & \footnotesize (d) spectrum \\
\end{tabular}
  \caption{Sample vs Isolated channels AF3 and AF4.}
  \label{fig:AF3vsAF4}
\end{figure}

\begin{figure}[H]
\begin{tabular}{>{\centering\arraybackslash}m{3.15cm}>{\centering\arraybackslash}m{3.15cm}>{\centering\arraybackslash}m{3.15cm}>{\centering\arraybackslash}m{3.15cm}}
 \makecell{\includegraphics[width=1\linewidth]{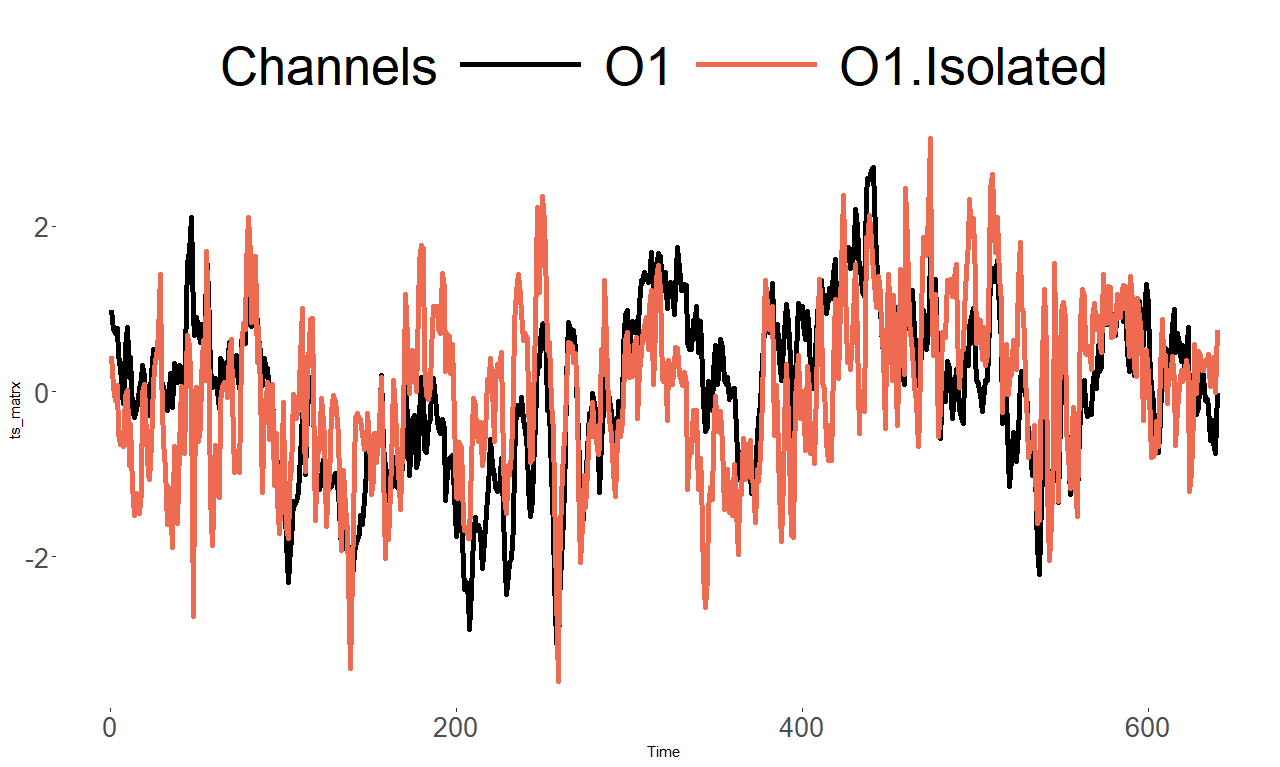} \\ \includegraphics[width=1\linewidth]{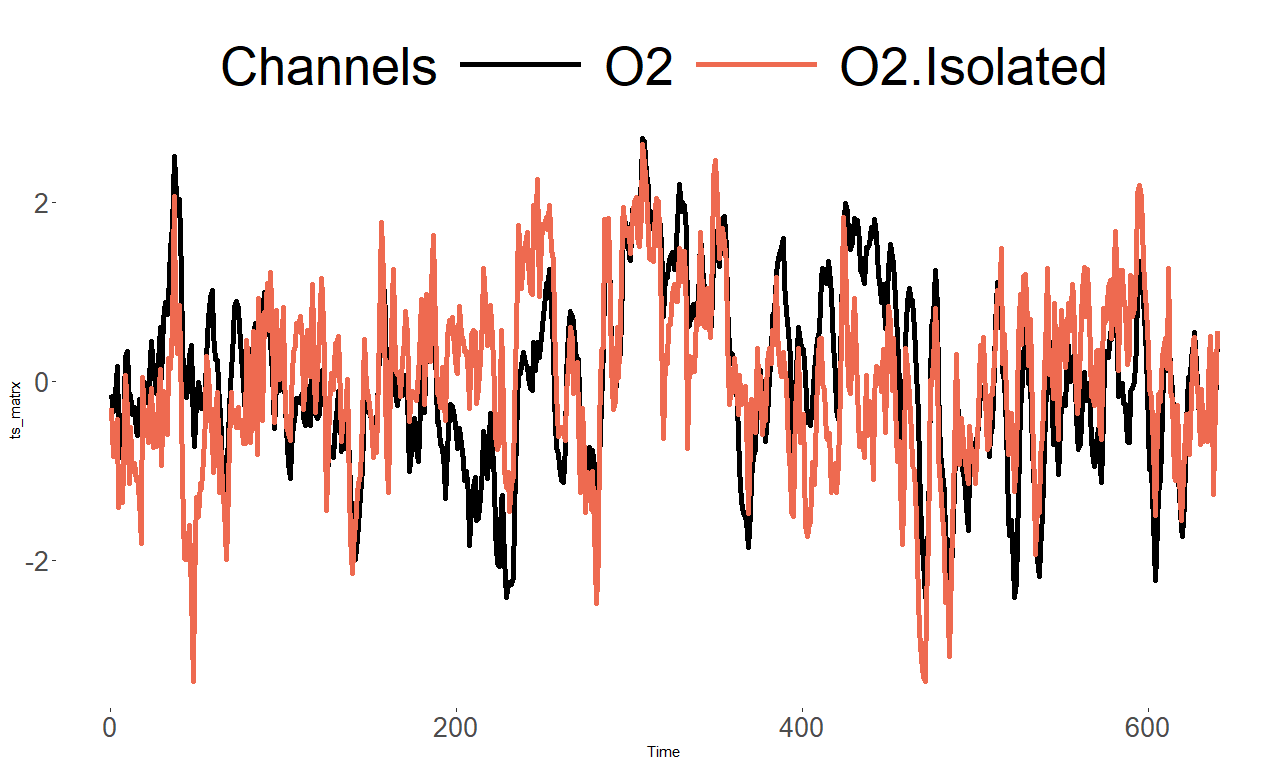}}
 & \makecell{\includegraphics[width=1\linewidth]{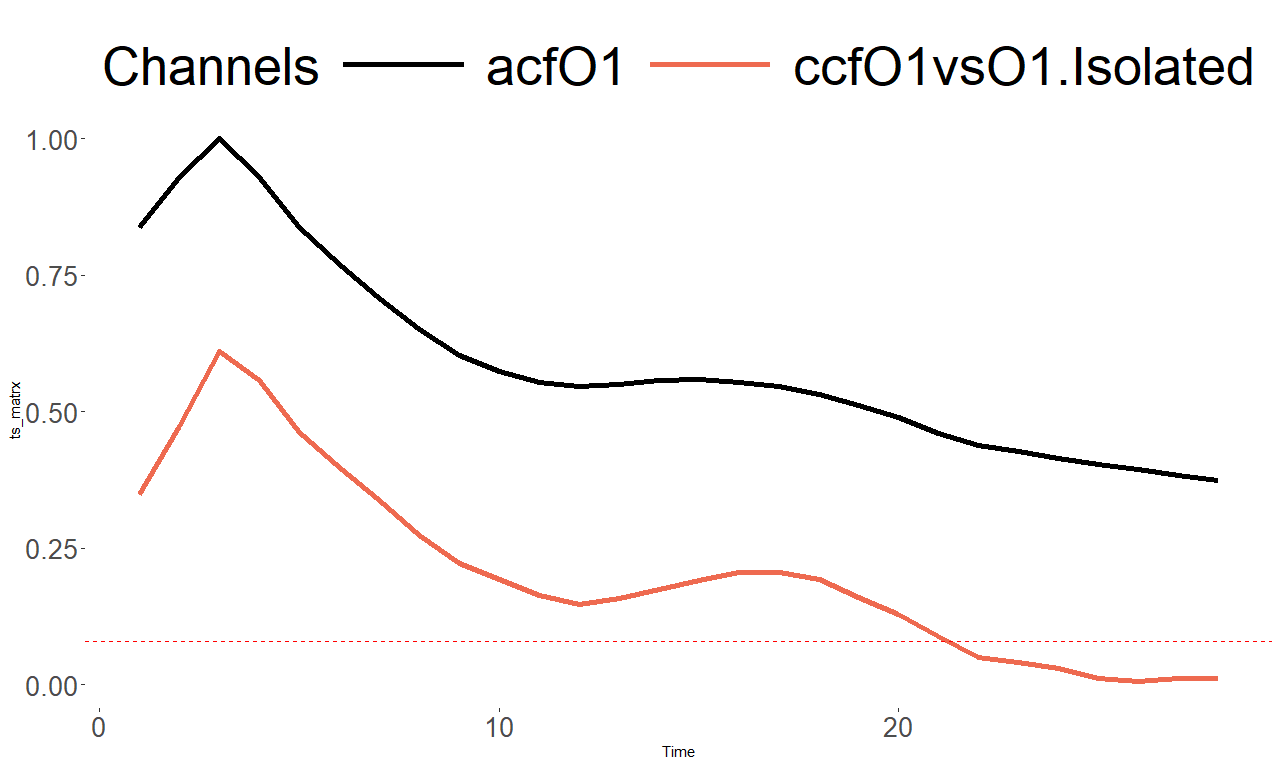} \\ \includegraphics[width=1\linewidth]{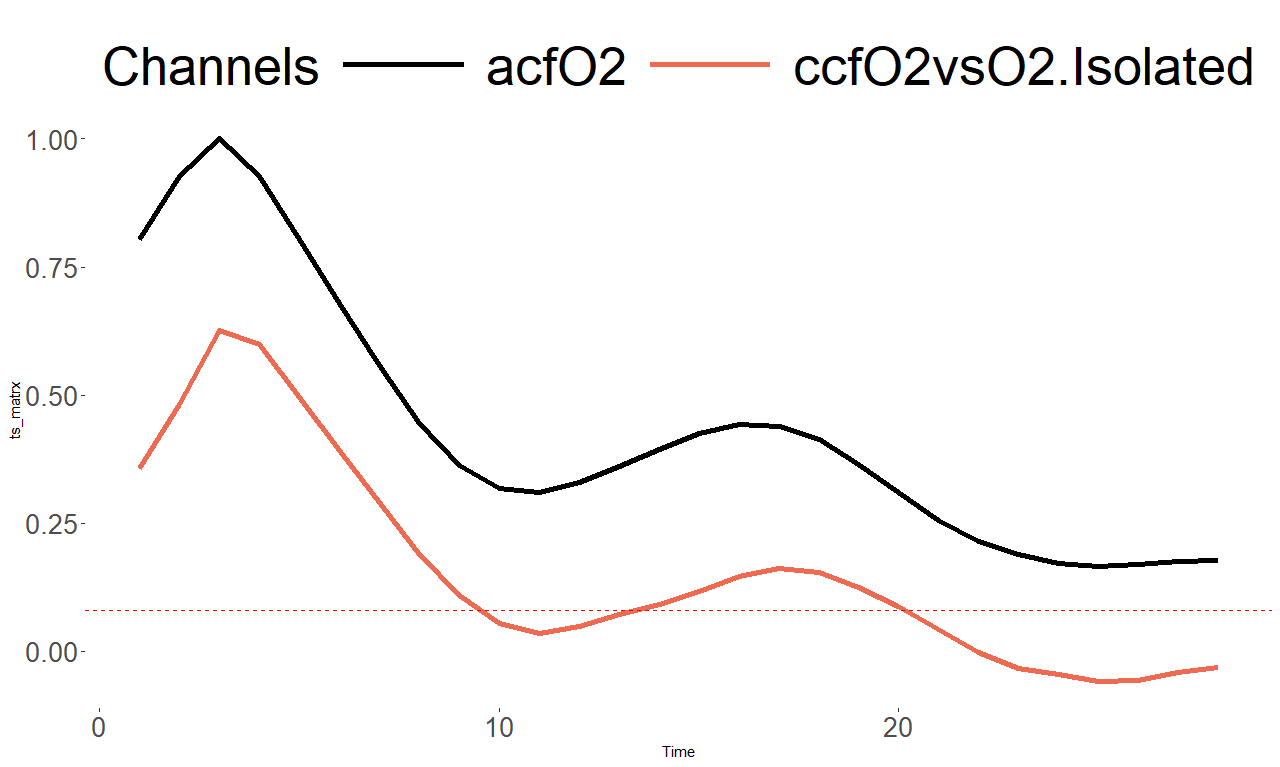}} & \makecell{\includegraphics[width=1\linewidth]{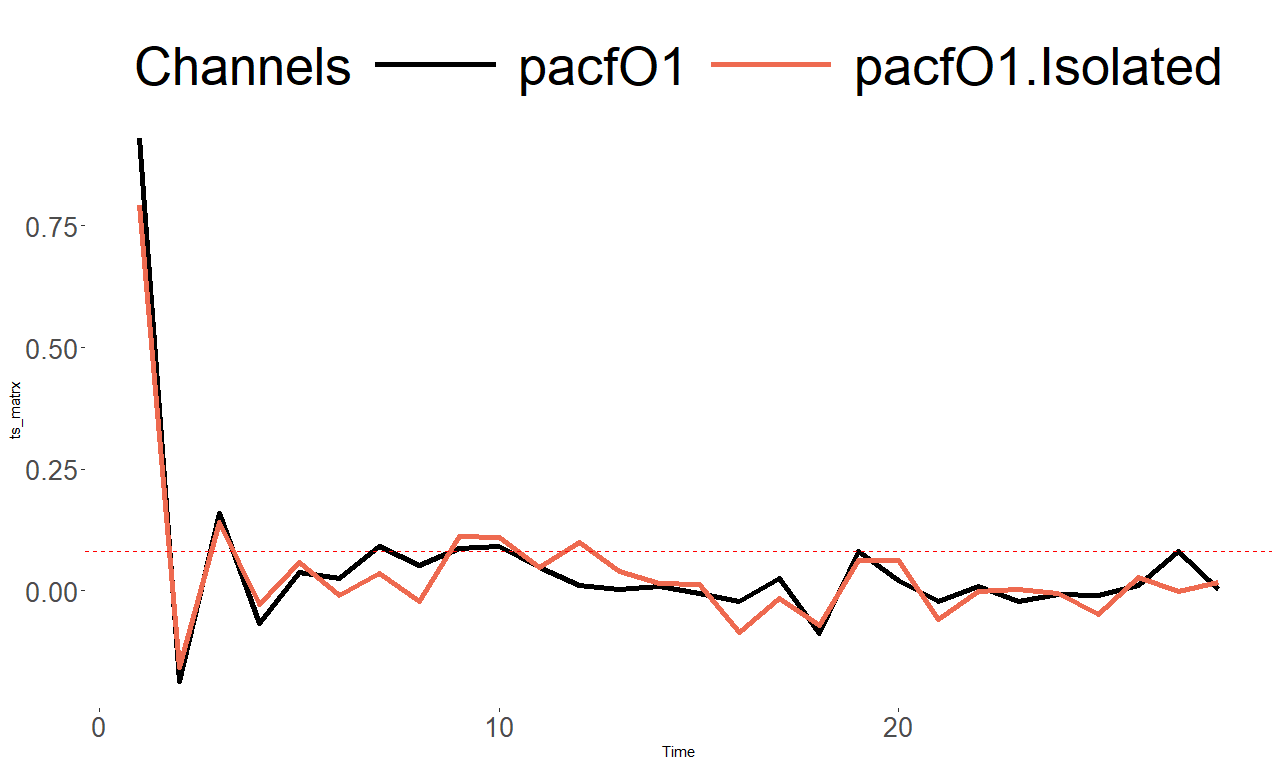} \\ \includegraphics[width=1\linewidth]{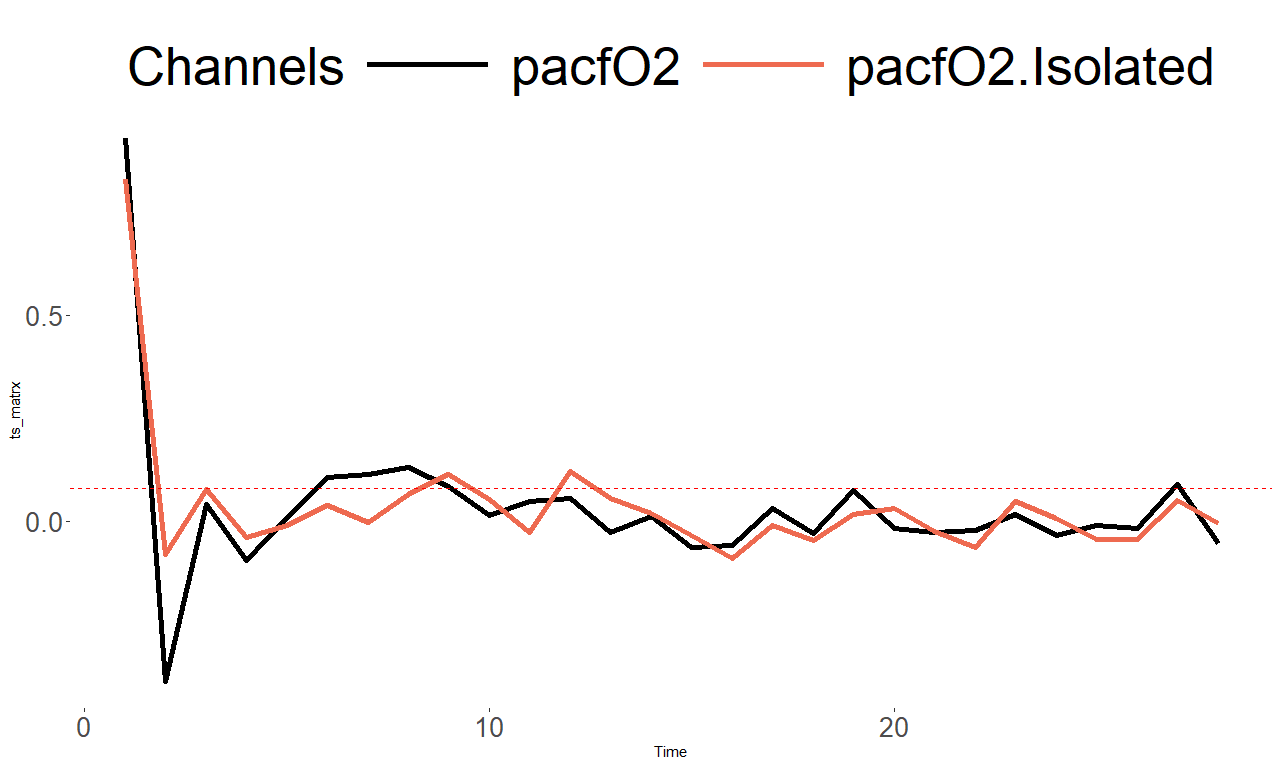}} & \makecell{\includegraphics[width=1\linewidth]{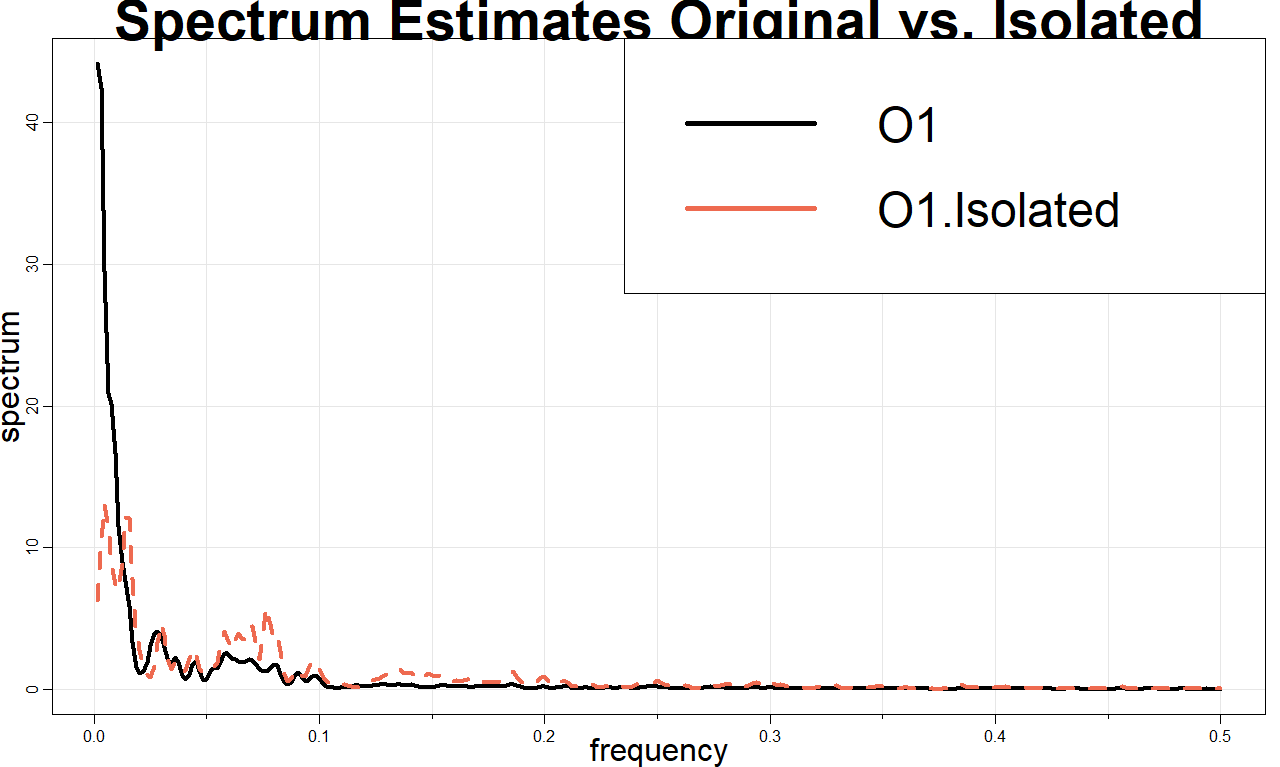} \\ \includegraphics[width=1\linewidth]{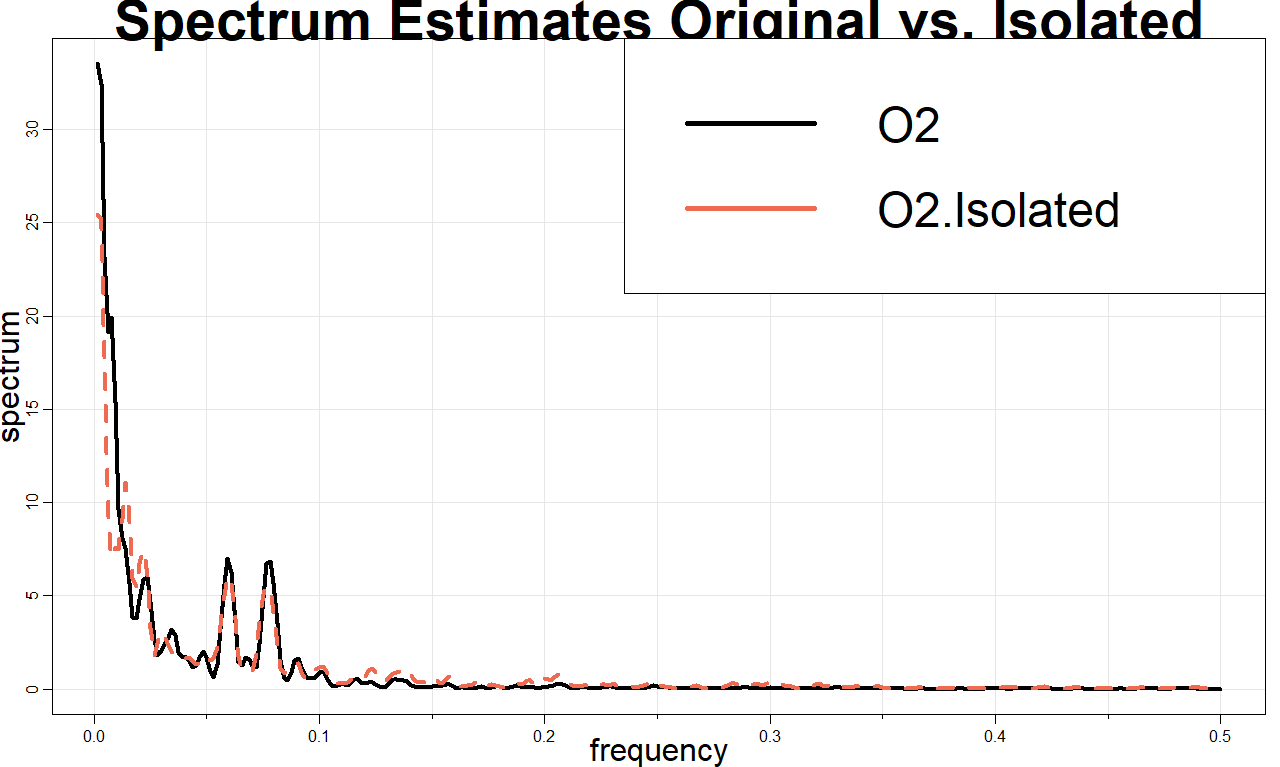}} \\
\footnotesize (a) trace & \footnotesize (b) acf & \footnotesize (c) pacf & \footnotesize (d) spectrum \\
\end{tabular}
\caption{Sample vs Isolated channels O1 and O2.}
\label{fig:O1vsO2}
\end{figure}

We compared the time-dependent properties before and after applying sDPCA to the channels of interest. As illustrated in Figures \ref{fig:C3vsC4}, \ref{fig:AF3vsAF4} and \ref{fig:O1vsO2}, although the temporal characteristics of autocorrelation are maintained, there is a noticeable trend toward less correlatedness at long lags, suggesting that the shared variance with other brain areas has been removed. This can be considered as an indication of isolation from influencing nodes.

The remainder of this section presents estimated connectivities among COI regarding 35 subjects pertaining to both motor execution (ME) and motor imagery (MI) tasks. The subjects performed a series of opening and closing (OC) gestures, including left-hand fist (LH), right-hand fist (RH), both fists' (BH), and both feet' (BF), for both the ME and MI tasks. The aim was to compare the brain activity associated with these tasks and gestures by identifying the patterns of connectivity and causal interactions amongst the selected channels. Each subject repeated the relevant task (i.e., task1: OC of LH/RH, task2: imagery OC of LH/RH, task3: OC of BH/BF, task4: imagery OC of BH/BF) 3 times (i.e., runs). Within these three task-oriented runs, they performed each motor movement/imagery motion (i.e., LH, RH, BH, BF) approximately 23 times in total. 

\begin{figure}[htbp]
    \centering
    \begin{subfigure}[b]{0.24\linewidth}
        \includegraphics[width=\linewidth]{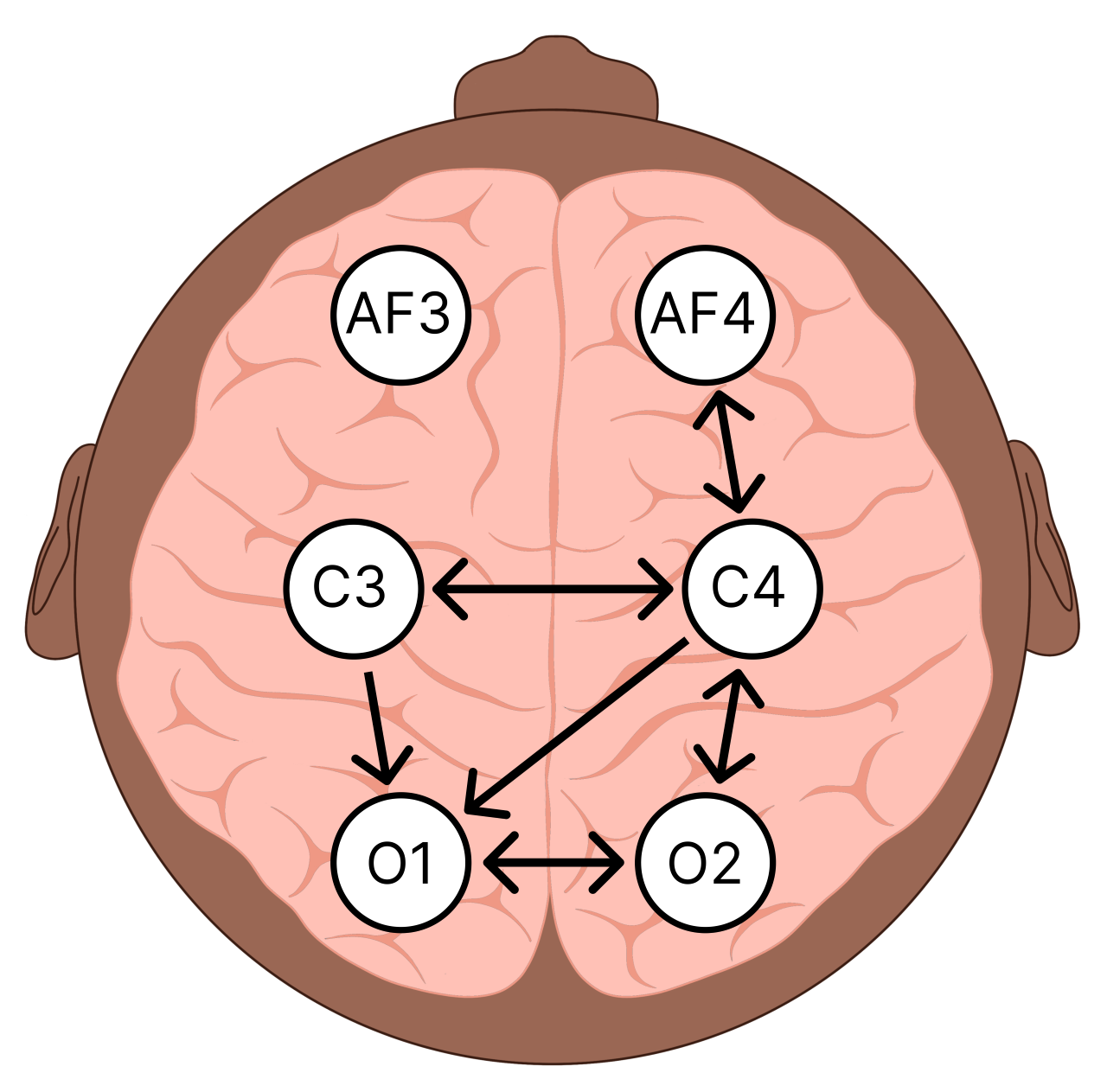}
        \caption{Left Fist (MI)}
        \label{fig:COIconnect_a}
    \end{subfigure}
    \begin{subfigure}[b]{0.24\linewidth}
        \includegraphics[width=\linewidth]{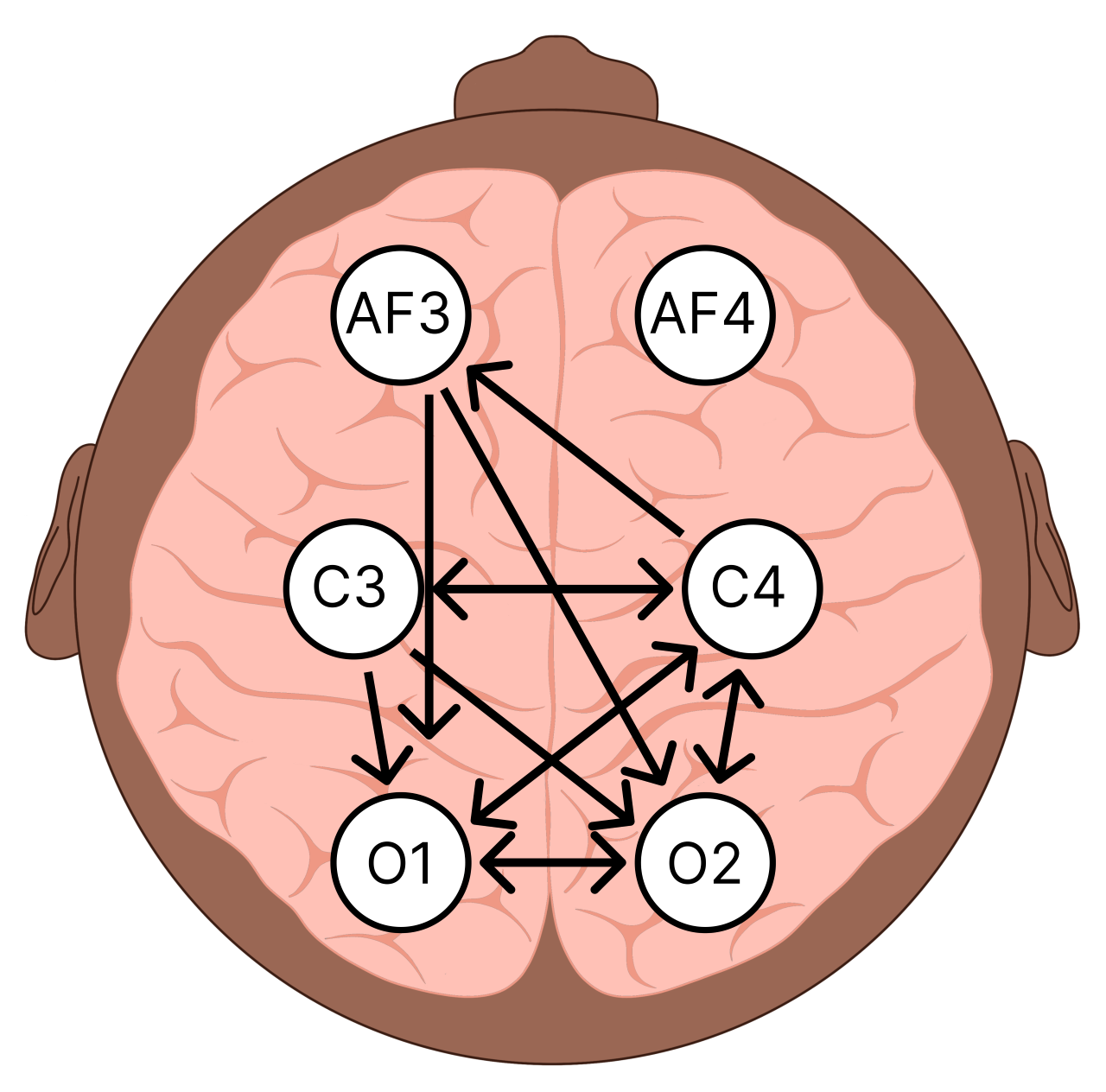}
        \caption{Right Fist (MI)}
        \label{fig:COIconnect_b}
    \end{subfigure}
    \begin{subfigure}[b]{0.24\linewidth}
        \includegraphics[width=\linewidth]{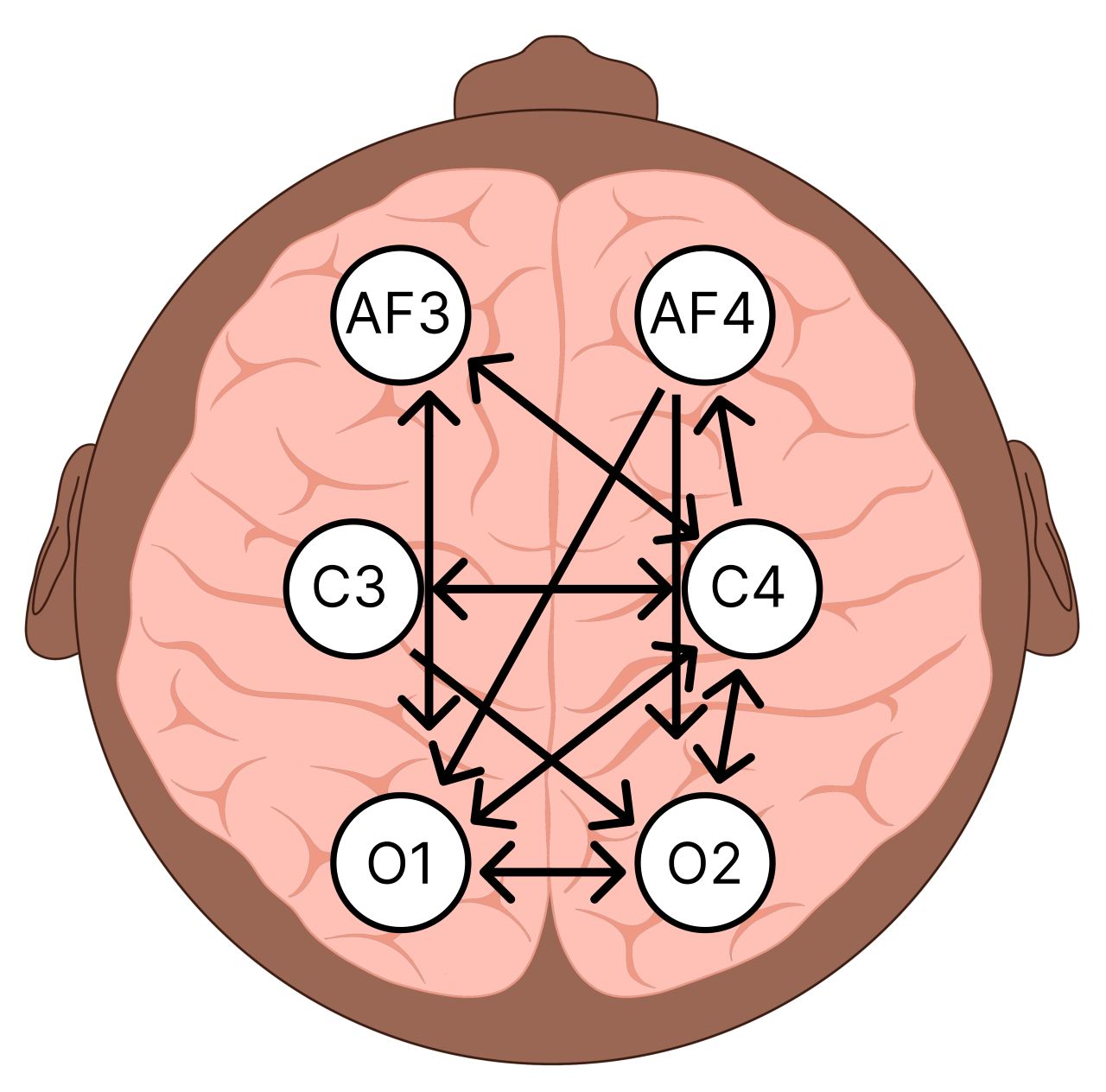}
        \caption{Both Fists (MI)}
        \label{fig:COIconnect_c}
    \end{subfigure}
    \begin{subfigure}[b]{0.24\linewidth}
        \includegraphics[width=\linewidth]{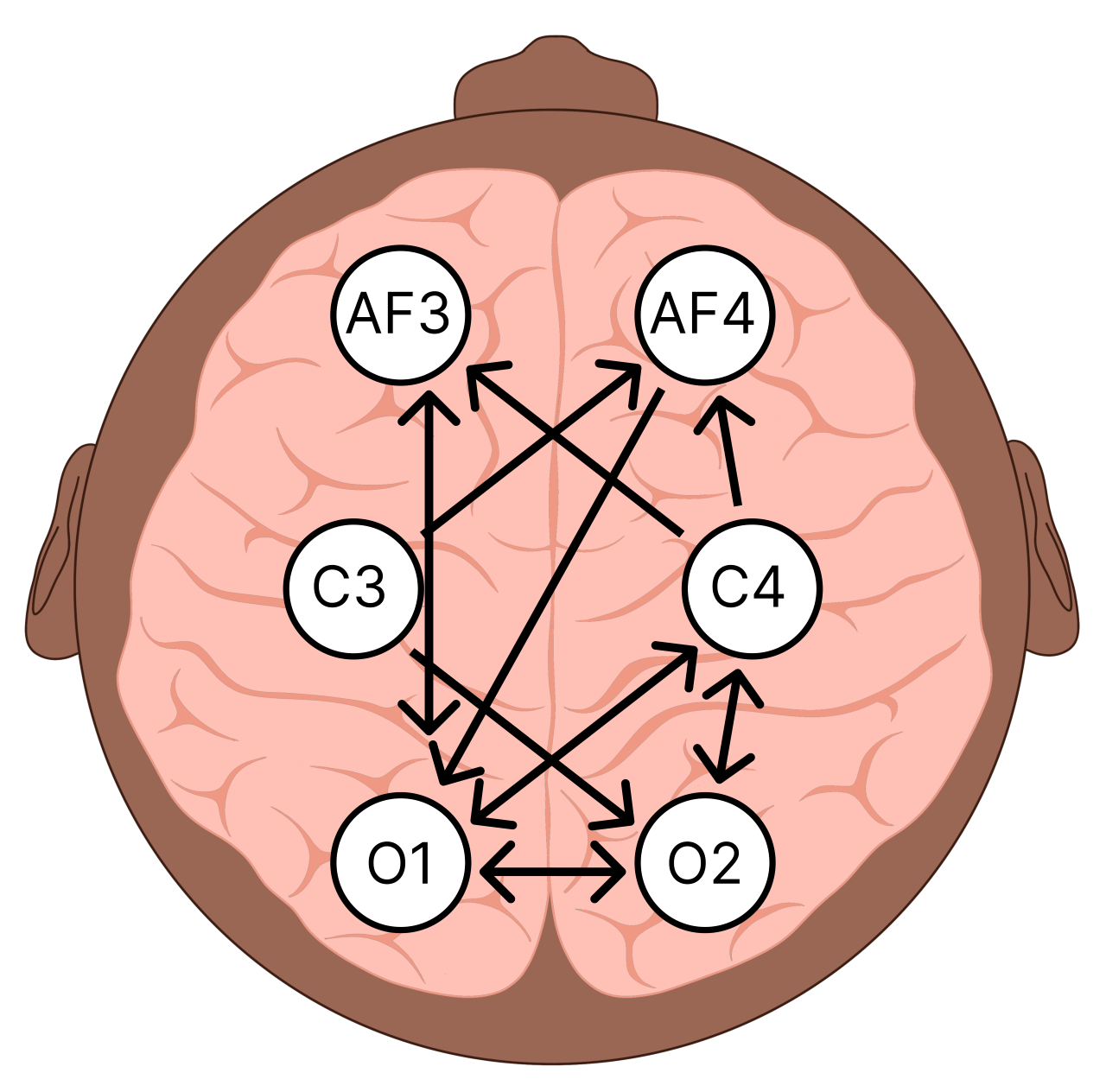}
        \caption{Both Feet (MI)}
        \label{fig:COIconnect_d}
    \end{subfigure}
    
    \vskip 1em  
    
    \begin{subfigure}[b]{0.24\linewidth}
        \includegraphics[width=\linewidth]{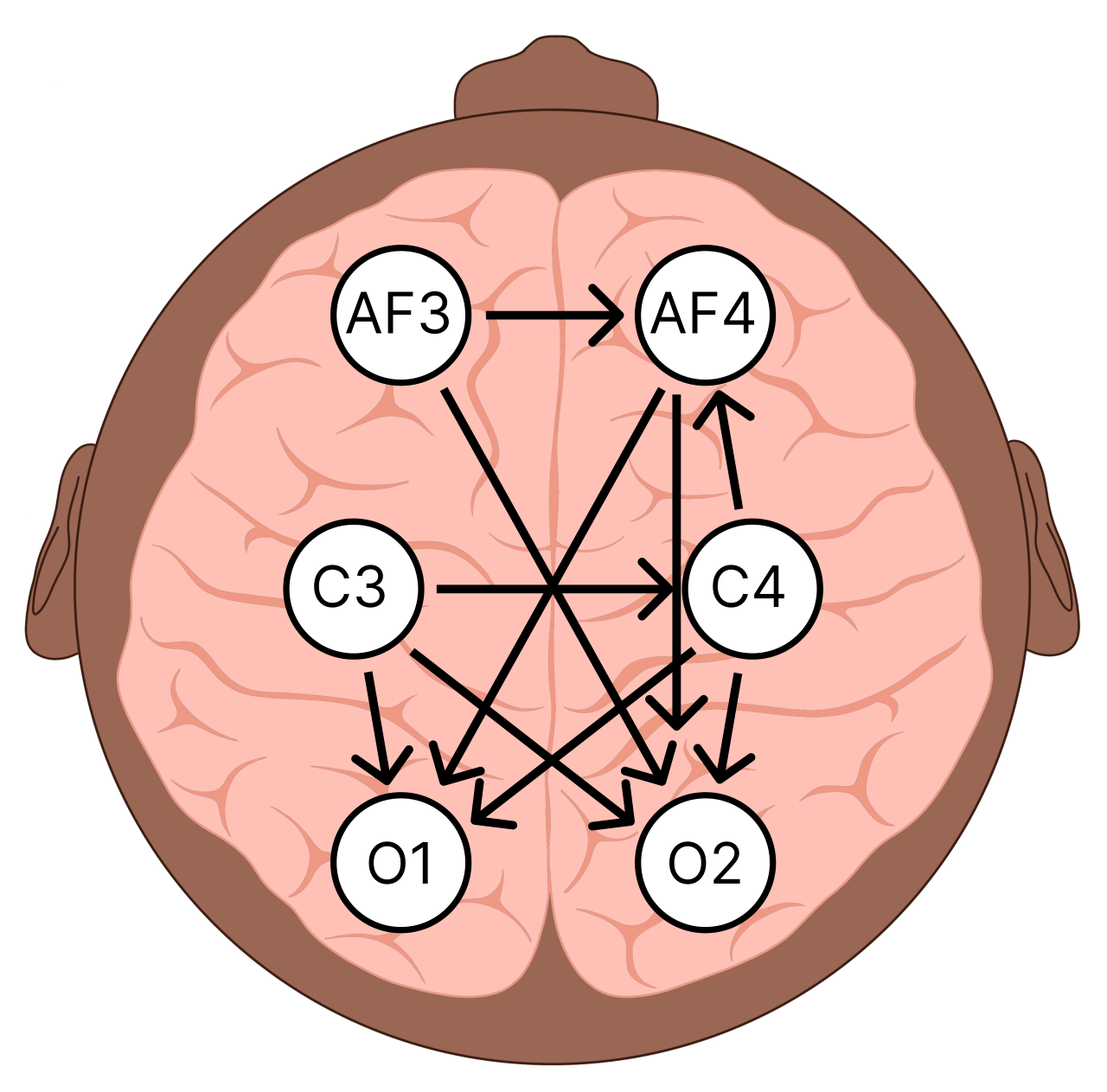}
        \caption{Left Fist (ME)}
        \label{fig:COIconnect_e}
    \end{subfigure}
    \begin{subfigure}[b]{0.24\linewidth}
        \includegraphics[width=\linewidth]{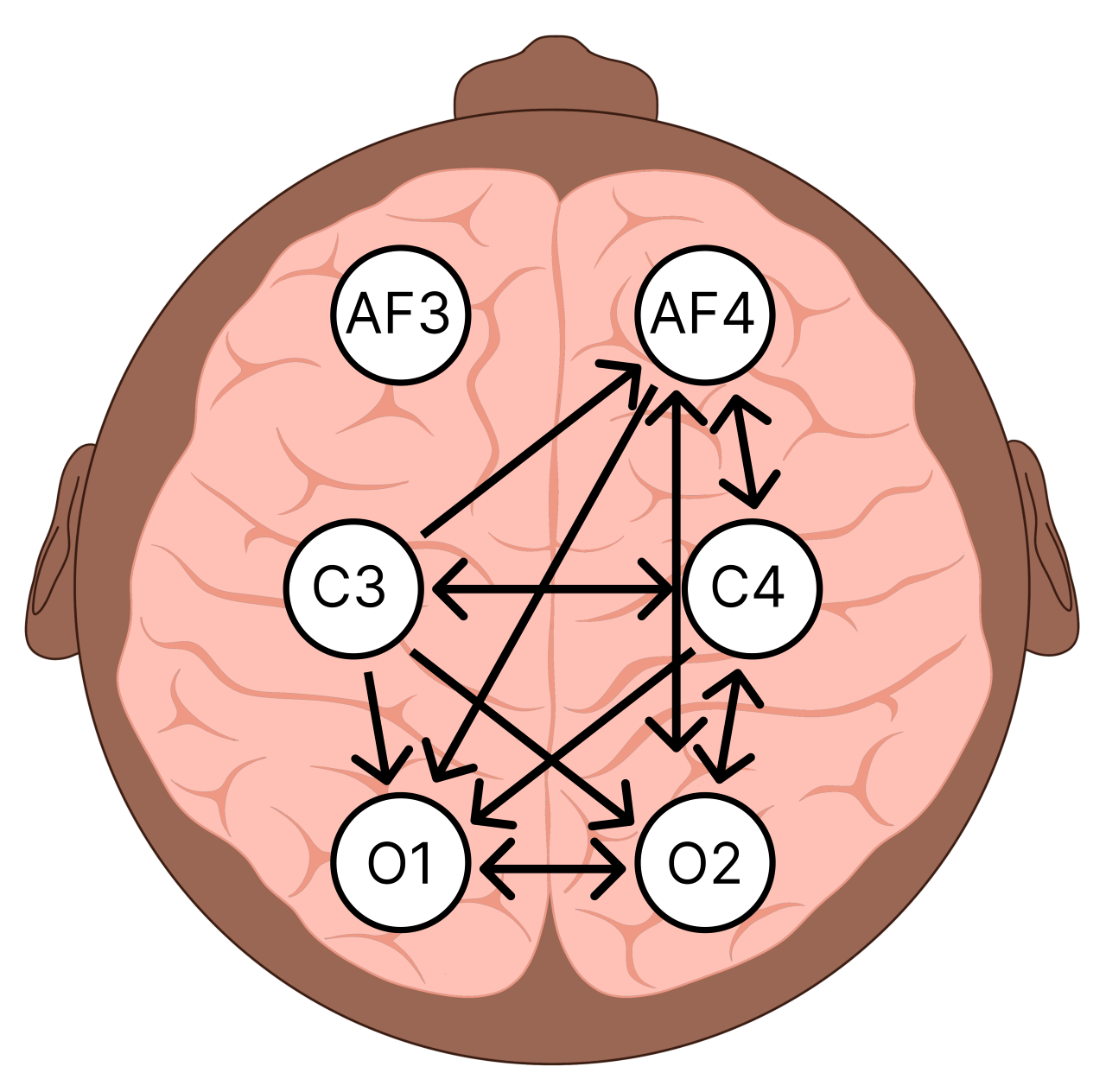}
        \caption{Right Fist (ME)}
        \label{fig:COIconnect_f}
    \end{subfigure}
    \begin{subfigure}[b]{0.24\linewidth}
        \includegraphics[width=\linewidth]{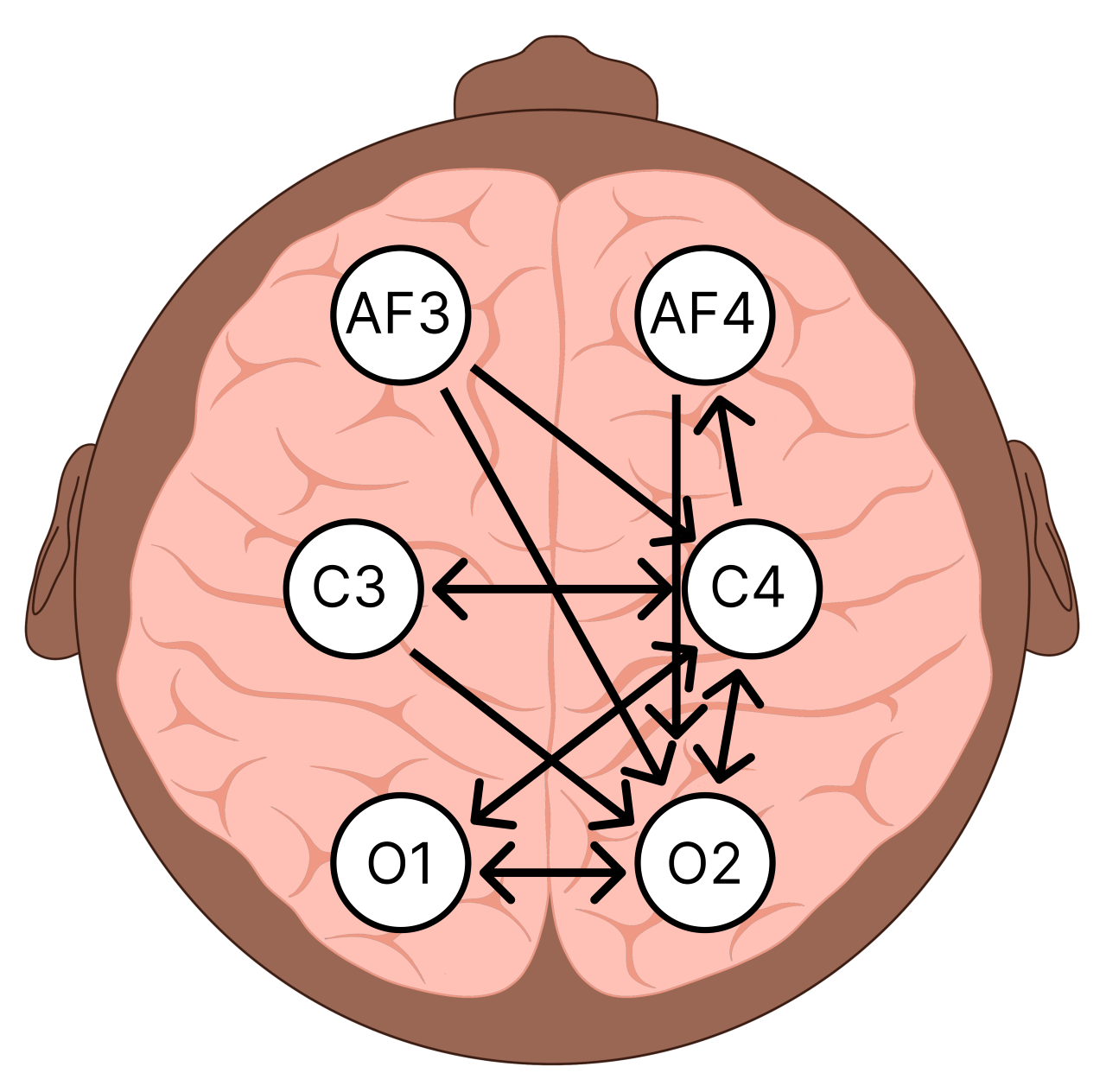}
        \caption{Both Fists (ME)}
        \label{fig:COIconnect_g}
    \end{subfigure}
    \begin{subfigure}[b]{0.24\linewidth}
        \includegraphics[width=\linewidth]{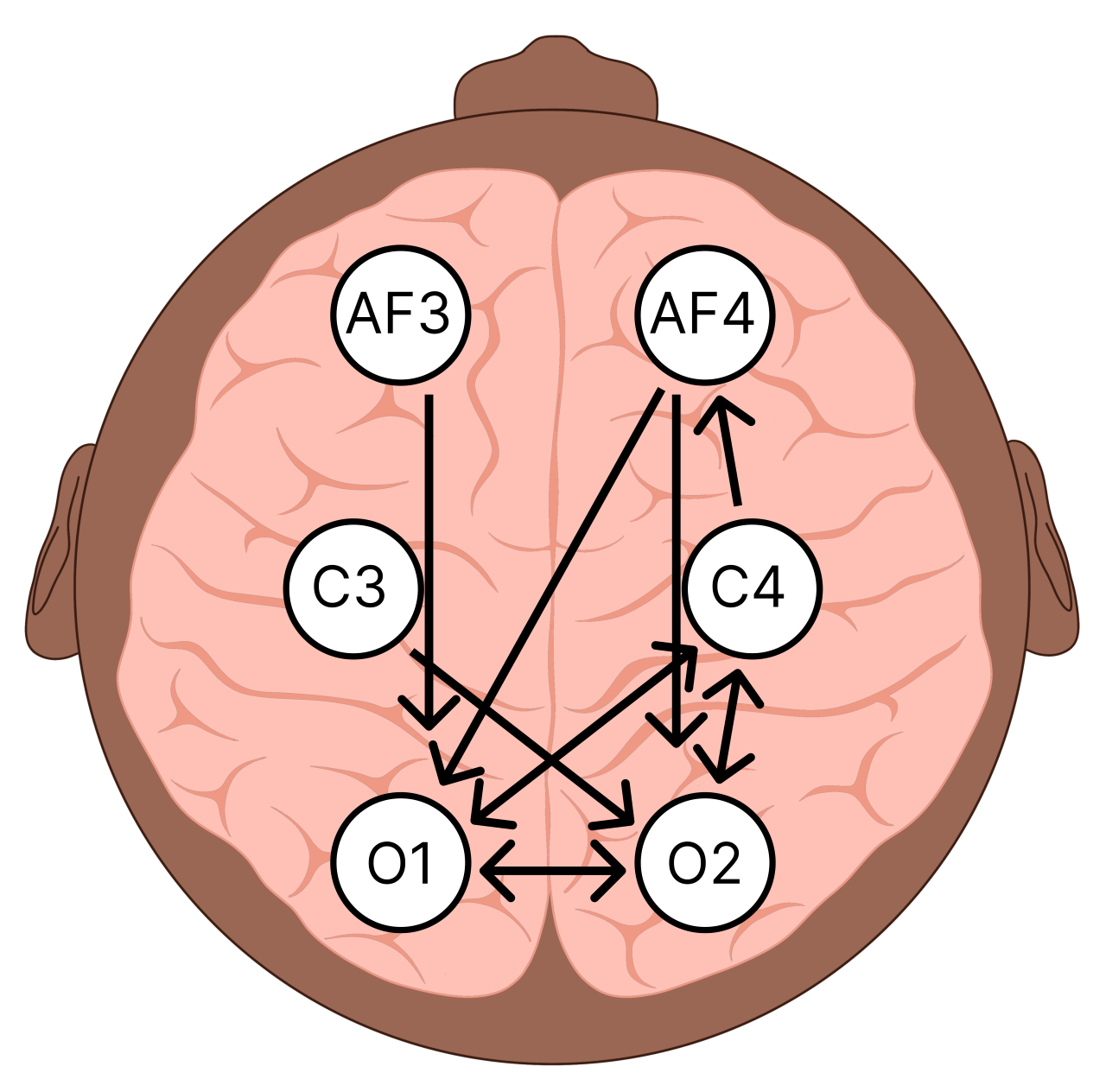}
        \caption{Both Feet (ME)}
        \label{fig:COIconnect_h}
    \end{subfigure}

    \caption{Connectivity under Motor Imagery (top row) vs.\ Motor Execution (bottom row). Each panel (a)--(h) shows connectivity for Left Fist, Right Fist, Both Fists, and Both Feet.}
    \label{fig:COIconnect}
\end{figure}

Each 64 dimensional epoch consists of about 4 seconds, that is, 640 observations. Initially, the relevant epochs were extracted from the dataset with the help of annotation information. In order to simulate a high-dimensional space for 64-dimensional EEG, we augmented the real dataset to include 170 additional signals to 64 channels by creating additional features. By increasing the dimension of each epoch, we also aimed to enrich the dataset to better capture the underlying neural dynamics. This was achieved through a series of data augmentation steps designed to capture various aspects of neural activity and connectivity. First, we generated hemispheric asymmetry channels by calculating the difference between pairs of electrodes located symmetrically across the left and right hemispheres (e.g., Fp1 - Fp2). Next, we computed anterior-posterior gradient channels by taking the difference between frontal and posterior electrode pairs (e.g., Fz - Pz). We also created lateral connectivity channels by finding differences between electrodes along the sides of the brain (e.g., FT7 - TP7). To examine central brain activities, we calculated midline activity channels by subtracting posterior midline electrodes from anterior ones (e.g., Fpz - Oz). Additionally, we computed regional average channels by averaging the signals from electrodes within specific brain regions—frontal, central, parietal, and occipital. We employed spatial derivative channels (Laplacian) by subtracting the average signal of neighboring electrodes from the signal at a central electrode (e.g., Cz minus the average of FCz, CPz, C1, and C2). This enhances the detection of localized neural activity by emphasizing differences from the surrounding area. Furthermore, we introduced product interactions between specific pairs of electrodes by multiplying their signals (e.g., F3 times P3). This captures interactions (i.e., possibly nonlinear) and joint activations between different brain regions, which might be significant for complex cognitive processes. Then, for each epoch, the high-dimensional data were reduced to COI dimensions by the proposed approach, and the estimated connectivities were recorded. From these results, the connectivities shared by at least 70\% of the subjects were determined and demonstrated in Figure \ref{fig:COIconnect}.

The GC results revealed distinct patterns of connectivity among the six channels of interest during different motor tasks, as illustrated in Figure \ref{fig:COIconnect}. When examining left-hand motor imagery, we observed that channel C3 (left motor cortex) G-causes channel C4 (right motor cortex) and channel O1 (left occipital cortex). Channel C4 also G-causes C3, AF4 (right pre-frontal cortex), O1, and O2 (right occipital cortex). This suggests that during the imagination of left-hand movement, there is significant bilateral interaction between motor cortices and connections to occipital regions involved in visual processing.

In contrast, during left-hand motor execution, the connectivity increased. Channel C3 G-causes C4, O1, and O2, indicating that ME engages both occipital cortices more than imagery does. The involvement of both visual regions during execution may reflect the integration of sensory feedback when performing physical movements. Similarly, for right-hand tasks, during motor imagery, channel C3 G-causes C4, O1, and O2, and channel C4 G-causes C3, AF3 (left pre-frontal cortex), O1, and O2. This pattern highlights strong bilateral connectivity and suggests that both hemispheres are actively engaged even during imagined movements.

When comparing motor imagery and execution for both fists, the connectivity patterns showed that channel C4 had extensive Granger causality to other channels during imagery, indicating the right motor cortex's central role in coordinating bilateral imagined movements. During execution, there was increased causality from AF3 and AF4 (pre-frontal cortices) to motor and occipital regions, suggesting greater involvement of executive functions and planning during actual movement. For both feet tasks, the connectivity patterns involve interactions among motor, pre-frontal, and occipital channels in both imagery and execution. Imagery and execution paths share influence from C4 onto multiple channels, indicating a common mechanism connecting the right motor cortex to pre-frontal and occipital regions. Findings suggest that BF movements share overlapping causal networks in both imagery and execution, but the directionality and intensity slightly differ.

Overall, the contrasts between motor imagery and execution reveal that actual movement involves more extensive and complex connectivity, particularly with increased involvement of pre-frontal regions during execution.

In conclusion, the proposed approach is an easy-to-implement method for examining the connectivity between a few nodes in a high-dimensional setting. By efficiently isolating the nodes of interest and removing the confounding effects of other channels, one may focus on specific neural interactions without the computational burden of analyzing the entire high-dimensional network.



\section{Discussion and Conclusion}

Our study aimed to enhance the usage of conventional Granger causality analysis in high-dimensional time-series networks, including application on EEG datasets. To this end, we investigated whether a two‑step strategy -sDPCA for dimensionality reduction followed by tests for Granger causality- can recover causal links between selected time series within a high-dimensional network. By projecting the activity of hundreds of background channels onto a handful of dynamic principal scores, the procedure lowers the parameter count, shortens computation time, and maintains the temporal information essential for EEG-like brain monitoring data. Simulated data confirmed that the proposed approach keeps false discoveries low and detects underlying connections if the background influence is not overwhelmingly strong. The EEG study provided a secondary line of evidence: after removing global confounding activity with sDPCA, the remaining channel pairs showed causal patterns corresponding to the known physiology of motor tasks. The results suggest that the approach captures spatial and temporal structure well enough to dissect meaningful interactions while avoiding the cost of fitting a full high‑dimensional VAR.

Several conclusions emerge. First, spectral decomposition supplies a compact set of scores that retain lag and frequency-based information under low-dimensional representation, allowing conventional F‑tests to remain valid. Second, the reduction step adds little overhead; its computational burden scales mainly with the Fourier transform used to build the spectral matrix, not the square of the channel count. Third, the isolation step mitigates the long lag serial correlation of signals, which helps to uncover connections between series more efficiently. Together, these points make the workflow practical when the goal is to focus on one or two lead-lag relations embedded in a large network.

The same findings also mark the limitations of the framework. When many channels share very strong drivers, relevant variance spreads across more dynamic components than can be kept without restoring noise, and detection power falls. Results also depend on a stable spectral density; abrupt shifts in frequency content undermine the eigen‑decomposition. Standard sDPCA relies on symmetric filters that are unusable in our context; replacing them with one‑sided causal filters prevents leakage of future information at the cost of lower frequency resolution. Finally, researchers interested in global connectivity still need sparse multivariate models. The proposed approach is not a universal replacement for full network modeling. When the scientific aim extends from a pair of nodes to the global network topology, sparse high‑dimensional modeling strategies remain necessary.

Future extensions could embed the current sDPCA setting within richer pipelines. One path is to follow the causal scores with sparse‑ or ridge‑regularised VAR estimation, while another is to expand the score vector with kernel or other nonlinear transformations so that interaction terms and threshold effects enter the GC testing. A sequential variant could let the one‑sided filter update online, allowing real‑time monitoring. The dimensionality‑reduction stage can be upgraded by forming a weighted blend of latent factors drawn from nonlinear ICA, autoencoder embeddings, and graph‑spectral filters; this composite factor set feeds into the same purge‑then‑GC workflow and equips the method to address a wider spectrum of confounding patterns.

In summary, sDPCA, followed by Granger causality testing, offers a practical compromise between full‑network modeling and bivariate analyses. It delivers reproducible inference on targeted node pairs while controlling for high‑dimensional background activity, provided that stationarity and moderate sparsity hold. These qualities make the approach useful for the toolkit for studying directed interactions in large‑scale time‑series data, such as EEG.

\begingroup
\bibliographystyle{chicago}
\bibliography{GCinHDN.bib}
\endgroup

\end{document}